\documentclass[12pt,english]{article}
\usepackage[utf8]{inputenc}
\usepackage{graphicx}
\usepackage{comment}
\usepackage{physics}
\usepackage{natbib}
\usepackage{microtype}
\usepackage{amsmath}
\fontfamily{cmss}\selectfont
\usepackage{libertine}
\usepackage{bm}
\usepackage{xcolor}

\newcommand{\1}[1]{{\color{black}#1}}
\newcommand{\2}[1]{{\color{black}#1}}
\newcommand{\3}[1]{{\color{black}#1}}
\newcommand*{\crosssymbol}{%
    \text{%
      \raise 1ex\hbox{%
        \rlap{\vrule height.2pt depth.2pt width .75ex}%
        \hbox to .75ex{\hss\vrule height .5ex depth 1ex\hss}%
      }%
    }%
}
\makeatletter
\usepackage[utf8]{inputenc}
\usepackage{amssymb}

\usepackage[font=normalsize,font=doublespacing,labelfont=bf,justification=raggedright]{caption}

\usepackage{geometry}
\title{\vspace{-2em} Not Quite Killing It: \\
 Black Hole Evaporation, Global Energy,\\ and De-Idealization.}

\author{Eugene Y. S. Chua \\ \textit{eugene.chuays@ntu.edu.sg} \\ \textit{Nanyang Technological University, Singapore}}

\date{Preprint of 11 February 2025. \\ Accepted for publication at the \textit{European Journal for Philosophy of Science}. \\Please cite published version when available.}
 
\begin{document}

\maketitle

\begin{abstract}
    \noindent A family of arguments for black hole evaporation relies on conservation laws, defined through symmetries represented by Killing vector fields which exist globally or asymptotically. However, these symmetries often rely on the idealizations of stationarity and asymptotic flatness, respectively. In non-stationary or non-asymptotically-flat spacetimes where realistic black holes evaporate, the requisite Killing fields typically do not exist. Can we `de-idealize' these idealizations, and subsequently the associated arguments for black hole evaporation? Here, I critically examine the strategy of using `approximately Killing' fields to de-idealize black hole spacetimes and approximately extend conservation laws to non-idealized cases. I argue that this approach encounters significant challenges, undermining the use of these idealizations to justify the evaporation of realistic -- rather than idealized -- black holes, and raising questions about the justified use of such idealizations. \\
    
\end{abstract}

\begingroup
\tableofcontents
\endgroup

\section{Introduction}

It is often said that black holes are where our best theory of matter, quantum mechanics, meet our best theory of spacetime, general relativity. One prominent way in which black hole physics connects both quantum mechanics and general relativity is through the study of \textit{black hole thermodynamics}. By adopting a semiclassical approximation, Hawking (1974/1975) first studied the effects of (classical) black hole horizons on a (quantum) vacuum field, and argued that black hole horizons can be understood as radiating at a certain temperature to observers at infinity. In other words, black hole horizons emit Hawking radiation. Beyond spawning other research programs, notably the quest to resolve the so-called information loss paradox,\footnote{See Belot et al (1999) for an excellent review of the information loss paradox.} Hawking radiation also motivates a broader project to study black holes as bona fide thermodynamic objects, stemming from Hawking's prediction that black holes should `evaporate' as a result of Hawking radiation. We should thus treat black holes as obeying thermodynamic laws just as ordinary matter does.\footnote{For a general rejoinder to black hole thermodynamics as more than a `formal analogy', see Dougherty \& Callender (2016). See also a response by Wallace (2018).} In Curiel's words, ``almost everyone agrees that black hole thermodynamics provides our best guide for clues to a successful theory of quantum gravity." (2019a, 27) The hope is that a closer investigation of black holes will unearth a more fundamental theory unifying both quantum mechanics and general relativity.\footnote{See e.g. Hawking (1977).} 

Since Hawking, black hole evaporation has come to be enshrined as a linchpin of black hole physics: a search for `black hole evaporation' in the physics literature will yield overwhelming consensus on its existence.\footnote{A notable outlier is Ellis (2015) which discusses similar concerns to the one I will discuss here, concerning the global conservation of energy (or lack thereof) and its role in motivating black hole evaporation.} There are, by now, many different approaches with which one can derive Hawking radiation and black hole evaporation, each with their theoretical baggage.\footnote{For a sampler of the variety of approaches, see e.g. Wall (2009), Wallace (2018).} \1{Studying the foundations of these various approaches have physical and philosophical upshots. Physically, examining the assumptions behind each approach can challenge us to improve and generalize the results from each approach, discern their limits, and better understand the physics of black holes as well as related questions, such as a clearer understanding of the nature of the information loss paradox.} \2{Philosophically, such scrutiny also raises interesting questions about the justified use of idealizations, whether de-idealizations are needed for such justification, and what happens if such de-idealizations cannot be found. While such questions have been discussed in the broader philosophy of science literature and in other contexts such as phase transitions, they have only been explored more seriously in the context of semiclassical gravity in recent years, by philosophers such as Gryb et al (2021) or Ryder (forthcoming).}

\2{In this vein, this paper will examine the foundations of one prominent family of arguments for black hole evaporation given Hawking radiation. Such arguments rely essentially on the global conservation of energy one way or another, and generally go something like this.\footnote{\2{That is, such arguments rely on global spacetime structure in a crucial way. These are contrasted against other approaches which do not obviously do so, such as those which employ an expected stress-energy tensor of the appropriate vacuum state on a black hole spacetime and considers the sort of vacuum polarization that happens near the horizon for such states. For the latter approach, see e.g. Birrell \& Davies (1982) and references therein. Thanks to an anonymous reviewer for suggesting that I distinguish these approaches.}} Since we observe black holes emitting energy via Hawking radiation, this energy must be coming from somewhere due to the conservation of energy.\footnote{Due to the myriad literature out there with differing terminologies, and the mass-energy equivalence in general relativity, I will use `mass', `mass-energy', and `energy' interchangeably in this paper.} If the spacetime we are studying is vacuum everywhere,\footnote{More precisely, the spacetime contains a quantum field in the vacuum state.} as in typical black hole spacetimes such as Schwarzschild spacetime, the only object that could lose energy as a result of the radiation is the object of study, the black hole. In other words, black holes lose mass (`evaporate') due to Hawking radiation, just as ordinary matter radiate and lose energy to their environment.}

In my view, however, this family of arguments for black hole evaporation rest on subtle foundational questions surrounding the nature of energy and idealizations in general relativity. Such arguments depend crucially on the assumption that there exists some \textit{globally} conserved energy, which in turn rely on the existence of appropriate global or asymptotic spacetime symmetries. Yet, it is not clear that these exist for non-idealized systems in general relativity, given a prominent view of idealization. 

\3{My worry for such arguments begins from the most naïve understanding of global energy conservation, in terms of global time-like Killing fields and stationarity. Given such an approach, an apparent dilemma for black hole evaporation arises: either the spacetime is time-independent, which guarantees the global conservation of energy par excellence but also ensures that no change over time -- and no evaporation -- occurs. Or the spacetime is time-dependent, so that evaporation over time is \textit{possible}. However, without time-independence, the naive approach seems to say that the relevant spacetime does not have a globally conserved energy after all.} 

What's interesting is not so much the dilemma itself -- after all, stationarity is rarely assumed (nothing interesting happens if nothing changes) -- but rather the sort of answers one might give to this dilemma. In practice, two prominent idealizations for introducing the global conservation of energy for \textit{non}-stationary systems are used instead (implicitly or explicitly). The first is the notion of \textit{quasi-stationarity}, and the second is \textit{asymptotic flatness}. Prominently, Hawking's original 1975 derivation explicitly used the notion of quasi-stationarity, while asymptotic flatness sees widespread use in black hole physics. At first glance, both allow one to introduce a global conservation law for energy in contexts where the metric is time-dependent. However, borrowing from recent discussions by Norton (2012, 2016) and Duerr (2019), I will argue that while both tools \textit{qua} idealizations can support inferences for black hole evaporation, they do not support the inferences to black hole evaporation in realistic -- `de-idealized' -- black holes. As I will argue, there is no clear way of `de-idealizing' from these idealizations in general relativity, to borrow from McMullin (1985), something that isn't appreciated enough in the philosophical literature.\footnote{See also Knuttila \& Morgan (2019) for a contemporary discussion of de-idealization.} 

In the philosophical sphere, there are lively debates surrounding the nature of gravitational energy \textit{in general}, though this debate has not, to my knowledge, been discussed explicitly in the context of black hole physics.\footnote{See the seminal Hoefer (2000), with responses by e.g. Pitts (2010), Read (2018) and De Haro (2022).} An outlier here is Maudlin (2017, 7), who briefly notes (without elaboration) that ``the cogency of Hawking’s 1975 argument that the black hole should lose mass and eventually evaporate is not completely evident, but that is a matter for another time". Wallace (2018) also provided some brief arguments in favor of black hole evaporation, though he does not contend with the problems I will raise here surrounding quasi-stationarity or asymptotic flatness. 

In what follows, I begin by reviewing relativistic black hole physics in \S2. In \S3, I sketch Hawking's argument for Hawking radiation and introduce the aforementioned family of arguments for how Hawking radiation, plus global conservation of energy, leads to black hole evaporation. In \S4, I introduce the naïve dilemma for black hole evaporation. In \S5 and \S6 I introduce quasi-stationarity and asymptotic flatness and argue that we are not yet justified in inferring that realistic black holes evaporate using these idealizations. Once we distinguish clearly the difference between approximation and idealization (following Norton's (2012) characterization), I argue that such arguments for black hole evaporation \textit{essentially} require these idealizations. I do so by arguing that there is no clear way yet to `de-idealize' these idealizations in terms of approximations. To do so is to provide a sense in which spacetimes \textit{approximately} have certain global or asymptotic symmetries; this, in turn, seem to require the notion of `approximate Killing fields'. However, I argue that, right now, there is no suitable notion of approximate Killing fields which can do the work required. Without a way to de-idealize these idealizations, we cannot yet conclude that arguments made with these idealizations apply to our world -- they seem to require these idealizations essentially. As such, these arguments for black hole evaporation which rely on global conservation of energy do not yet provide us with justification for thinking that realistic black holes, too, evaporate. This points to a direction for future theoretical work -- in rigorously demonstrating that such a de-idealization can be obtained. 

In \S6.5 I discuss how this problem can be quarantined from most applications of general relativity: those applications, such as the predictions of gravitational lensing, the perihelion precession of Mercury, or accretion and black hole growth due to infalling matter, do \textit{not} essentially require the problematic idealized properties discussed here. I also briefly consider the more controversial case of gravitational radiation. To that end, my argument does not `infect' all of general relativity. Finally, I want to emphasize that I am \textit{not} ruling out the possibility of black hole evaporation \textit{per se}; rather, this paper only seeks to evaluate one family of arguments -- those relying on global spacetime structure to motivate the use of global conservation of energy to infer that black holes evaporate -- and to point out how one prominent view for justifying idealizations appears to fail for those arguments here. The use of some idealizations in general relativity do not appear to have a `clean' de-idealization procedure, which raises further questions about how idealizations are to be justified in this context. As I mentioned earlier, there are many other arguments for black hole evaporation and Hawking radiation, which I cannot cover in a single paper (perhaps not even a book). Furthermore, there are other theories of idealization and de-idealization, some of which might be more permissive, which might vindicate and justify black hole evaporation despite the worries I have presented here. (In fact, I am currently working on precisely one such account of idealization and de-idealization.)  

\section{An overview of relativistic black hole physics}

I begin by specifying some key concepts. Since arguments for black hole evaporation typically take place in the semiclassical regime, gravity is understood classically. The arena of discourse is thus general relativity. 

Our discussion begins from the metric tensor $g_{\alpha \beta}$ (henceforth simply the `metric'), which defines a spacetime of interest and constrains the behavior of matter on said spacetime via the Einstein Field Equations:\footnote{More precisely a spacetime is given by the pair ($M, g_{\alpha\beta}$) where $M$ is a manifold. Here I will speak of the metric and spacetime interchangeably. Nothing turns on this difference. Furthermore, to simplify presentation, I will use natural units such that $c = G = \hbar = k = 1$.}  
\begin{equation}
    R_{\alpha \beta} - \frac{1}{2}Rg_{\alpha \beta} + \Lambda g_{\alpha \beta} = 8\pi T_{\alpha \beta}
\end{equation}
where $R_{\alpha \beta}$ is the Ricci tensor, $R$ is the Ricci scalar, $\Lambda$ is the cosmological constant (which may or may not vanish), and $T_{\alpha \beta}$ is the stress-energy tensor encoding the behavior of matter in spacetime.

In general, energy is always \textit{locally} conserved in general relativity. Along any worldline $\chi$, the covariant derivative of $T_{\alpha \beta}$ vanishes:

\begin{equation}
    \nabla_\chi T_{\alpha \beta} = 0
\end{equation}
In other words, momentum and energy are conserved given infinitesimal displacements along any worldline, as one would expect from classical physics. 

However, as is well-known, this does not generally entail \textit{global} conservation of energy.\footnote{See Maudlin, Okon \& Sudarsky (2020) for an excellent discussion.} The fact that energy is conserved along any observer's worldline does not allow us to say that energy is conserved for the \textit{entire} spacetime. 

Famously, Noether (1918) showed that every differentiable symmetry of the action of a physical system is associated with some conserved current satisfying a continuity equation, and thus a corresponding conservation law. In the context of general relativity, these symmetries are represented by Killing vector fields (or simply Killing fields) which generate isometries (trajectories along which the metric is constant). A Killing vector $\boldsymbol{\xi}$ represent an infinitesimal displacement along which the Lie-derivative ($\mathsterling$) of the metric vanishes: 
\begin{equation}
    \mathsterling_{\boldsymbol{\xi}} g_{\alpha \beta} = 0
\end{equation}
This demand leads naturally to the result that $\boldsymbol{\xi}$ satisfies Killing's equation, where $\nabla$ is the covariant derivative: 
\begin{equation}
    \nabla_{\nu}\xi_\mu + \nabla_{\mu}\xi_\nu = 0
\end{equation}
With Killing's equation and the geodesic equation, where $\textbf{p}$ is the tangent vector to any arbitrary geodesic,
\begin{equation}
    \nabla_\textbf{p} \textbf{p} = 0
\end{equation}
we can derive the following theorem.\footnote{See Misner, Thorne \& Wheeler (1973, 651) for discussion.} In any spacetime geometry endowed with a symmetry described by a Killing field $\boldsymbol{\xi}$, motion along any geodesic leaves the scalar product of the tangent vector $\textbf{p}$ with the Killing vector $\boldsymbol{\xi}$ constant \1{(taken with respect to the metric $g_{\mu\nu}$):}
\begin{equation}
    \textbf{p} \cdot \boldsymbol{\xi} = g_{\mu\nu} p^{\mu}\xi^{\nu} = constant
\end{equation}
This allows us to describe a \textit{globally} conserved quantity on a spacetime, i.e., a quantity conserved everywhere in spacetime.\footnote{See Hawking \& Ellis (1973, 61 -- 63), Misner, Thorne \& Wheeler (1973, \S25.2), Carroll (2019, 120), or Maudlin, Okon \& Sudarsky (2020, \S2.4) for discussion.} For example, space-like translational symmetries are what allow us to make sense of the global conservation of linear momentum, while the space-like rotational symmetries let us define the global conservation of angular momentum. Likewise, the time-like translational symmetry, represented by the existence of Killing fields along the time-like coordinate, is associated with the global conservation of energy. \3{That is, it suffices for global conservation of energy that there exists a global time-like Killing field, i.e., an isometry of the metric along the time-like direction.}\footnote{See, also, Brown (2022) who details the equal footing of both Noether's theorem and its converse. This suggests a strict two-way relationship between conservation laws and symmetries, though some technical caveats apply.} \3{As I'll discuss in much more detail later in \S6, it also suffices for global conservation of energy if the spacetime is endowed with appropriate \textit{asymptotic} symmetries, specifically, the existence of asymptotic time-like Killing fields associated with asymptotic time-like translational symmetry `at infinity', even if the appropriate global symmetries do not obtain. However, the existence of either global or asymptotic asymmetries is crucial: without such symmetries, we cannot define a global conservation law. If a spacetime does not have the requisite symmetries, there can be no global conservation laws for that spacetime. This will be crucial for our discussions later.}

Now we are in a position to consider the metrics of black holes, as well as their symmetries and associated conservation laws. The most common metrics associated with black holes (and the ones used in the proof of Hawking radiation to be discussed later) \textit{do} have some degree of symmetry, allowing us to define global conservation laws on spacetimes describing such black holes. As we will see, these laws are essentially tied to a key property of these black hole metrics: notably, it requires their \textit{time-independence}. 

For instance, the Schwarzschild metric describing the vacuum asymptotically flat exterior of a non-rotating uncharged spherically symmetric black hole in Schwarzschild coordinates ($t$, $r$, $\theta$, $\phi$) is given by:\footnote{I will discuss asymptotic flatness in much more detail in \S6.}

\begin{equation}
    ds^2 = -(1 - \frac{2M}{r})dt^2 + (1 - \frac{2M}{r})^{-1}dr^2 + r^2(d\theta^2 + \sin^2\theta d\phi^2) 
\end{equation}

\noindent where $M$ is the mass parameter and $2M$ is the Schwarzschild radius which determines the event horizon. Importantly, $M$ is constant in time here, i.e. a time-independent parameter (Schwarzschild 1916, 2). The metric components are independent of $t$ and $\phi$, and coordinate transformations reveal two more spatial rotational symmetries. Together the Schwarzschild metric has 4 associated Killing fields -- one time-like and three space-like -- resulting in global conservation of both energy and angular momentum in the usual spatial directions. \1{Importantly, the Schwarzschild metric is not only independent of the time coordinate $t$ but also lacks any cross terms mixing $dt$ with spatial coordinates. This implies the existence of a timelike Killing vector field that is hypersurface-orthogonal. Therefore, the Schwarzschild spacetime is \textit{static}, meaning it is time-independent and has no rotational or twisting effects.}

The Kerr metric, which describes the vacuum asymptotically flat exterior of a \textit{rotating} uncharged \textit{axially} symmetric black hole, is given in Boyer-Lindquist coordinates ($t$, $r$, $\theta$, $\phi$) by:

\begin{equation}\
\begin{aligned}
    ds^2 ={} & - \frac{\Delta - a^2\sin^2\theta}{\rho^2}dt^2 - 2a\frac{2Mr\sin^2\theta}{\rho^2}dtd\phi \\
    & + \frac{\rho^2}{\Delta}dr^2  + \frac{(r^2 + a^2)^2 - a^2\Delta \sin^2\theta}{\rho^2}\sin^2\theta d\phi^2 + \rho^2d\theta^2
\end{aligned}
\end{equation}

\noindent where $\Delta \equiv r^2 - 2Mr + a^2$, $\rho^2 \equiv r^2 + a^2 \cos^2 \theta$, $J$ is the angular momentum parameter and $M$ is again the mass parameter. The event horizons occur where $\Delta = 0$. Once more, by inspection, we can see that the Kerr metric is independent of the time-like $t$ as well as the space-like $\phi$. However, since the black hole is rotating, there is a privileged axis of rotation which rules out the two other Killing fields associated with the spherically symmetric Schwarzschild black hole. So we only have two Killing fields, one time-like and one space-like.\footnote{There is also a Killing \textit{tensor} field, though I will not discuss it in this context.} As a result, we have global conservation laws for energy and angular momentum (in one direction). \1{The Kerr metric is independent of the time coordinate $t$ but includes cross terms like $dt\,d\phi$ due to the rotation of the black hole. This introduces a mixing of time and space coordinates, resulting in a timelike Killing vector field that is \textit{not} hypersurface-orthogonal. Consequently, the Kerr spacetime is \textit{stationary but not static} -- it is time-independent but includes rotational effects that prevent the spacetime from being static.}\footnote{A generalized family of stationary metrics is the Kerr-Newman family of metrics, which also allows one to discuss charged black holes.}  

\section{From Hawking radiation to black hole evaporation}

The key idea for Hawking radiation is that we can consider how quantum matter fields behave near a collapsing star as the latter forms a black hole and event horizon, and how this behavior appears to observers during `late times' at infinity, i.e. after the black hole has settled into a \textit{stationary} or \textit{static} state. By comparing quantum fields at past and future infinity in a black hole spacetime, Hawking (1975) argued that the stationary black hole horizon -- a global spacetime structure -- can be interpreted as emitting radiation, and hence having a temperature $T$, proportional to its surface gravity $\kappa$: 
\begin{equation}
    T = \frac{\kappa}{2\pi}
\end{equation}
As mentioned, this has been derived in a variety of ways.\footnote{See e.g. Hawking (1974/1975), Wald (2001) or Carroll (2019).} Hawking's original calculations considered only spherically symmetric collapse, though, as Wald (2001, 12) notes, the effect obtains for any arbitrary gravitational collapse into a black hole. Note, however, that a common assumption remains despite the myriad of generalizations available nowadays: the radiating black hole is assumed \textit{stationary}. (Wald 2001, 12). We'll return to this in \S4 and \S5.

While Hawking's derivations establish that black holes can be thought as emitting some amount of energy via Hawking radiation, and that this radiation \textit{can} be interpreted as the temperature of the black hole, these derivations do not yet amount to an conclusive argument that the black hole \textit{really} has a temperature. As Wallace notes, 
\begin{quote}
these derivations in of themselves do not suffice to establish that Hawking radiation is fully analogous to ordinary thermal radiation, because they imply nothing about whether a radiating black hole ultimately decreases in mass and, thus, surface area. (2018, 11)    
\end{quote}
Ordinary thermodynamic objects lose energy and/or mass when radiating or losing energy to their surroundings. We have shown that global spacetime structures like black holes radiate energy to their surroundings, but do they lose energy or mass as a result? In other words, do they evaporate? As Hawking himself noted in his original derivation for Hawking radiation and black hole evaporation:
\begin{quote}
the particle creation [i.e. Hawking radiation] is really a global process and is not localised in the collapse: an observer falling through the event horizon would not see an infinite number of particles coming out from the collapsing body. Because it is a \textit{non-local} process, it is probably not reasonable to expect
to be able to form a local energy-momentum tensor to describe the back-reaction of the particle creation on the metric. (1975, 216)
\end{quote}
That is, the sort of energy conservation we should employ cannot be the usual local conservation of energy (eq. 2). As Wallace also observes:
\begin{quote}
    given that there is no robust local definition of gravitational energy and, relatedly, no robust way to understand total energy as a sum of local energies, we cannot simply appeal to a local conservation law to conclude that radiating black holes evaporate. (2018, 11)
\end{quote}
Because there is no robust local notion of gravitational energy,\footnote{The most prominent proposal is the pseudotensor approach. For a recent proponent of this approach, see Read (2020). For (what I think are successful) rejoinders and partial rejoinders, see Duerr (2021) and De Haro (2022) respectively.} we cannot appeal to local conservation laws for energy to support the argument for black hole evaporation. 

To my knowledge, though, many physicists rely implicitly or explicitly on \textit{some} notion of energy conservation when discussing the back-reaction of Hawking radiation on black holes. A generic answer is that black holes lose energy when Hawking radiation occurs, because Hawking radiation carries energy away to infinity. The assumption here is that energy is conserved somehow, so that any change in energy must be compensated by a corresponding change elsewhere. Hawking (1975) himself phrases the reasoning as follows:
\begin{quote}
    [Hawking radiation] will give positive energy flux out across the event horizon or, \textit{equivalently}, a negative energy flux in across the event horizon. [...] This negative energy flux corresponding to the outgoing positive energy flux will cause the area of the event horizon to decrease and so the black hole will not, in fact, be in a stationary state. (1975, 219, emphasis mine)
\end{quote}
The `equivalence' here amounts to some reasoning relying on conservation of energy. Hawking (1976) uses this reasoning more explicitly: 
\begin{quote}
    Because this radiation carries away energy, the black holes must presumably lose mass and eventually disappear. (Hawking 1976, 2461)
\end{quote}
Wald (2001) says the same:
\begin{quote}
    Conservation of energy requires that an isolated black hole must lose mass in order to compensate for the energy radiated to infinity by the Hawking process. (2001, 16)
\end{quote}
\noindent Carlip (2014, 20) simply equates the change of mass of a black hole to the power radiated by the black hole:\footnote{\2{Carlip is not the first to do so -- see e.g. Page (1976).}}
\begin{equation}
    \frac{dM}{dt} = -\epsilon \sigma T^4 A
\end{equation}
where $\epsilon$ is the emissivity parameter, $\sigma$ is the Stefan-Boltzmann constant, $A$ is the area of the black hole, and $T$ is its temperature. The right-hand side is simply the Stefan-Boltzmann law for the total power radiated by a black body over some surface area, but the claim that this \textit{equates} to the change in mass requires conservation reasoning. 

How are we to interpret all these claims about the use of conservation of energy in generating black hole evaporation, in light of the inability to employ a local conservation law of energy for such contexts? It seems that the next natural option is to appeal to some \textit{global} conservation law for energy. As I will argue in the next section, this approach, taken naïvely, leads quickly to contradiction. 


\section{A naïve dilemma for black hole evaporation}

\3{Since the typical argument for Hawking radiation and black hole evaporation is semiclassical, the operative assumption is that the classical rules of general relativity hold. This should include the rules for when a global conservation law exists, e.g., when a spacetime is stationary (see \S2). Given these rules, one might consider a naïve first step for justifying the use of global conservation of energy and arguing for black hole evaporation: understand it in terms of global time-like Killing fields and stationarity. After all, stationarity suffices for global energy conservation, and the aforementioned black hole metrics precisely possess the requisite time-like Killing fields.

Unfortunately, the story is not so simple, and this naïve perspective quickly leads to contradiction. Simply put, situations where evaporation is expected to occur are precisely those where there is no global time-like Killing field. On the contrary, situations where there is such a global symmetry are those where evaporation is impossible. So from this naïve perspective, the argument for black hole evaporation cannot take off. Granted, no one holds this naïve perspective. However, it is instructive for us to examine this argument briefly, for other tools -- to be discussed in later sections -- can be seen as ways to avoid this problem.}

Let me reformulate the worry in the form of a dilemma \2{(see also Ryder (forthcoming) for discussion of a paradox with a similar flavor, which he calls the black hole idealization paradox)}. For any spacetime in which we want to argue for the occurrence of black hole evaporation, we begin by noting that any such spacetime is either static/stationary, or not. In other words, the metric for that spacetime is either time-independent or not. 

On the one hand, if the spacetime is static or stationary, then it does have the appropriate global Killing field structure, as we have already seen from \S2. For example, the Kerr and Schwarzschild metrics are time-independent and do have global time-like Killing fields defined on their associated spacetimes. Indeed, as I mentioned above, the black hole is typically taken to settle into a stationary state when it is radiating Hawking radiation. Yet, we have also seen that static/stationary metrics are precisely those that do not change over time. Since they do not change over time, the black hole being described by said metric does not change over time either. This means said black hole with a static/stationary metric does \textit{not} evaporate, for evaporation entails a change of mass over time (as Carlip (2014) puts it explicitly above), rendering mass a time-dependent parameter. Since mass is a parameter featured in the metric, its time-dependence entails the time-dependence of the metric. Black hole evaporation cannot occur for a black hole described by a static/stationary metric.  

On the other hand, if a spacetime (i.e. the metric describing it) is \textit{not} static or stationary, then of course the parameters of the black hole in that spacetime are allowed to be time-dependent. They \textit{can} change over time and therefore we can describe the evaporation of a black hole with such a metric. For instance, mass can be a parameter that depends on the time-like coordinate. A simple example discussed by Wallace (2018) is the Vaidya spacetime, where the (retarded time) metric is:
\begin{equation}
    ds^2 = -(1 - \frac{2m(u)}{r})du^2 - 2dudr + r^2(d\theta^2 + \sin^2\theta d\phi^2)
\end{equation}
which looks a lot like the Schwarzschild metric with a time-dependent mass parameter. However, in these cases we do \textit{not} have a global time-like Killing field for the spacetime in question, since the actual metric is \textit{time-dependent}. Spacetimes which allow for evaporation are precisely spacetimes that do \textit{not} have the requisite global symmetries. \3{From the naïve perspective,} there is no global conservation law to rely on since there are no global symmetries, and hence we cannot yet assume some loss of mass or energy in these cases as compensation for the positive energy flux radiated to infinity by Hawking radiation. Put another way, spacetimes which \textit{can} accommodate evaporation are precisely those where we have \textit{no justification} yet for thinking that evaporation occurs.

\3{Thus, if we understand global energy conservation naïvely in terms of stationarity, we lack justification for believing in the occurrence of black hole evaporation on both horns of the dilemma. Again, to emphasize, I discuss this dilemma not because I think this naïve perspective is correct, but because I think it is a useful starting point for discussing other ways for motivating global energy conservation and justifying black hole evaporation. 

Given the failure of the naïve approach to global energy conservation, the rest of the paper will evaluate two distinct strategies for recovering global conservation of energy. These two strategies appeal to two different idealizations, mapping onto the two ways -- globally or asymptotically -- in which one can define global conservation of energy. On the one hand (\S5), one can accept the absence of global symmetries but appeal to approximate global symmetries via the idealization of \textit{quasi-stationarity} and approximate global time-like Killing fields. On the other hand (\S6), one can appeal instead to (approximate) asymptotic symmetries at infinity via a different idealization, \textit{asymptotic flatness}. As I'll argue in detail, both idealizations run into trouble when it comes to `de-idealization', and hence in justifying their use for modeling realistic systems.}

\section{Quasi-stationarity: the impossible process}

\subsection{Quasi-stationarity}

An immediate reply to the first horn of the naïve dilemma above is to point out that stationary systems are not physically interesting, since they never change over time. But the second horn bites back: non-stationary spacetimes do not have a well-defined conserved energy -- what then? 

Hawking himself was aware of this problem.  Evaporation can only occur for a \textit{non-stationary} black hole. However, the black hole was assumed to be \textit{stationary} in the derivation of Hawking radiation. How can we reconcile the two? In response, Hawking suggested we employ the assumption of \textit{quasi-stationarity}:
\begin{quote}
    This negative energy flux will cause the area of the event horizon to decrease and so the black hole will not, in fact, be in a stationary state. However, as long as the mass of the black hole is large compared to the Planck mass $10^{-5}$ g, the rate of evolution of the black hole will be very slow compared to the characteristic time for light to cross the Schwarzschild radius. Thus, it is a reasonable approximation to describe the black hole by a sequence of stationary solutions and to calculate the rate of particle emission in each solution. (Hawking 1975, 219)
\end{quote}
In short, quasi-stationarity is the claim that a black hole changing over time can be approximated with a sequence of stationary (or static) solutions. Though this assumption is used almost everywhere in black hole physics as Page (2005, 10) notes, it is not always spelled out explicitly.\footnote{Hawking (1975) and Frolov \& Page (1993) are some exceptions.} Typically it is masked by describing the black hole as `slowly evolving' (Zurek \& Thorne 1985), or by discussing black hole dynamics while using a static or stationary metric like the Schwarzschild or Kerr (i.e. \textit{non}-dynamical) metrics (Abramovicz \& Fragile 2013). 

Here's how an explicit rendition of Hawking's defense might look like. We start by accepting that the black hole (and spacetime) in question is actually time-\textit{dependent} and \textit{non-stationary}. However, because its mass is changing so slowly in time when it has a large mass relative to the Planck scale, we can assume that this time-dependent black hole approximates a certain idealization -- a sequence of time-independent stationary black holes that can be said to be \emph{quasi-stationary}. For each such stationary metric, there is global conservation of energy associated with that solution. So we can derive Hawking radiation in this stationary regime, and using conservation reasoning here, conclude that the energy of the black hole must decrease. This is where we calculate the aforementioned `negative energy flux'. 

This decrease, of course, cannot happen within \textit{any} stage of the idealized sequence of stationary black holes which form the quasi-stationary process; as Hawking notes, no change can happen by definition. Instead, we say that this decrease applies to the \textit{actual} time-dependent black hole where evaporation \textit{can} occur. We do this by perturbing the mass parameter of the Schwarzschild metric, bringing it from one stationary state to another in a sequence of stationary states. This is the quasi-stationary process representing black hole evaporation. In short, we perform the conservation reasoning in the idealized stationary regime but apply the results of this reasoning to the realistic target system being modelled.

\subsection{Quasi-staticity: the impossible process}

The above use of quasi-stationarity is reminiscent of one commonly found in equilibrium thermodynamics. There, as is well-known, equilibrium states are those whose thermodynamic parameters (e.g. volume, pressure, temperature, etc.) do not change with time. That is, equilibrium states are time-independent states. Yet, thermodynamics frequently make use of \textit{quasi-static} processes consisting of sequences of these equilibrium states.\footnote{See, for instance, Carathéodory (1909, 366).} For instance, the famous Carnot cycle describing a system interacting with two heat baths with differing temperatures can be described on a pressure-volume diagram, where each point is a state with unchanging pressure and volume. (see Figure 1) 

\begin{figure}[ht]
\includegraphics[scale = 0.4]{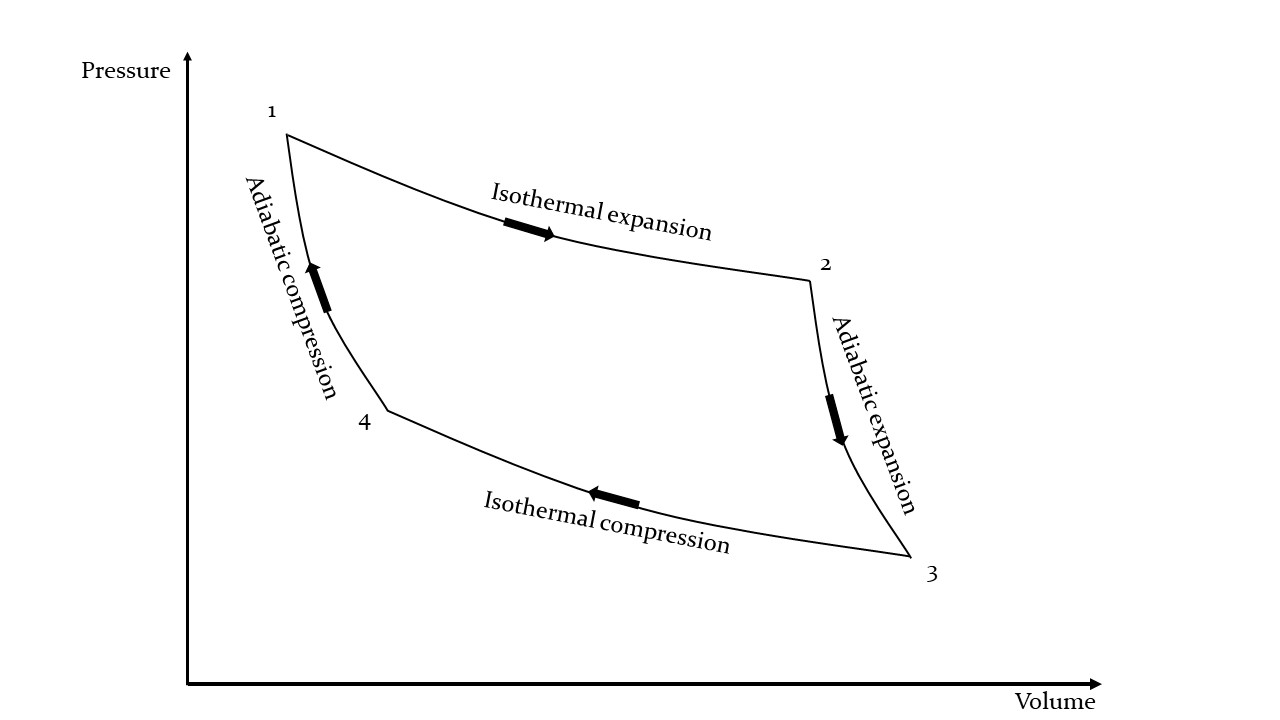}
\centering
\caption{The typical Carnot cycle.}
\end{figure}

In these cases, we are modeling target systems that are \textit{really} time-dependent using these idealized quasi-static processes. Typical thermodynamic objects change over time: our cup of coffee cool down -- and our mug of beer warms up -- over time. Nevertheless, if changes to these objects are slow and small enough, the system can be approximated with quasi-static processes, such that we can treat them as effectively time-independent at any point (or short interval) of time. We can then perform all thermodynamic calculations in terms of this quasi-static idealization, while keeping in mind that these calculations really apply to that underlying time-dependent process. This appears to vindicate the derivation for black hole evaporation -- we assume quasi-stationarity for conservation reasoning, and then apply the fruits of this reasoning to the actual system, which is just what we do in classical thermodynamics. 

Unfortunately, quasi-static processes in classical thermodynamics are not conceptually innocent. By inspecting why they work for classical thermodynamics, we can see why the analogous use of quasi-stationarity in the case of black hole evaporation does not. 

As Norton (2016) recently argued, quasi-static processes, too, come with their own internal tensions. Quasi-static processes are constituted by `sequences' of equilibrium states, each of which is approximated by an actual physical system at some time. These sequences are meant to be sequences of states \textit{in time} -- curves on thermodynamic state space (e.g. Figure 1) are parametrized by a time-like parameter. Furthermore, typical quasi-static processes describe variations in equilibrium states. Figure 1, for instance, includes curves with varying volume and pressure, although each point on these curves represents an equilibrium state. In other words, we are supposed to envision changes to these equilibrium states over time. Yet, equilibrium states do not change with time \textit{by definition}. So on the face of it, quasi-static processes change over time, but also do not change over time -- an outright contradiction.

The correct way to interpret quasi-static processes, as Norton argues, is not to take them to be actual processes with exactly those properties discussed above. Rather, actual processes only ever approximate quasi-static processes. The distinction between approximation and idealization is a subtle one, though one prominent distinction was recently introduced by Norton (2012). An approximation is characterized by being ``an inexact description of a target system", while an idealization is ``a real or fictitious system, distinct from the target system, some of whose properties provide an inexact description of some aspects of the target system." (Norton 2012, 209) Norton gives one simple example of this distinction: that of a body of unit mass falling in a weakly resisting medium. Its velocity $v$ at time $t$ is:
\begin{equation}
    \frac{dv}{dt} = g - kv
\end{equation}
where $g$ is the acceleration due to gravity and $k$ is a coefficient representing friction. When falling from rest at $v = t = 0$, its velocity can be expanded as:
\begin{equation}
    v(t) = \frac{g}{k}(1 - e^{-kt}) = gt - \frac{gkt^2}{2} - \frac{gk^2t^3}{6} - ...
\end{equation}
When there is low friction (i.e. $k$ is small), the fall of the ball is almost exactly described by: 
\begin{equation}
    v(t) = gt
\end{equation}

\noindent In terms of Norton's distinction, we can say that (14) inexactly describes the ball's descent. However, we can, in Norton's terms, `promote' this approximation to an idealization by having (14) directly refer to a fictitious system, that of a ball falling in a vacuum such that $k = 0$. Hence, (14) exactly describes such a system though the system need not exist (i.e. (14) is here an idealization), while it only provides an inexact description when the system is not in vacuum (i.e. (14) is here an approximation). 

Using this distinction, we may understand quasi-static processes, taken literally, as idealizations -- they can only be fictitious systems since they are contradictory in nature. These idealizations may \textit{approximate} realistic, actual, systems, of course, but crucially this means that they need not share all properties with the actual system -- they are \textit{inexact} descriptions. For instance, we think that the target systems being approximated by quasi-static processes are still time-dependent ones. As Norton (2016) shows, sets of these time-dependent processes may come arbitrarily close to quasi-static processes by having e.g. vanishingly small driving forces, but no actual process ever has all the exact properties of quasi-static processes (with exactly vanishing driving forces at each point of time). We cannot simply `take the limit' and let the driving force actually go to zero, for we have already seen how that leads to contradiction. As Norton puts it: 
\begin{quote}
    We have a sequence of [non-equilibrium] processes, each of which is slowed by diminishing the driving forces. Each process carries the property of completing a change, while requiring ever more time to do it. The limit of this property is the property of completing a change. The limit approached by the processes themselves, however, is no process at all. It is merely a static set of states in equilibrium that no longer carry the limit property of completing the change. (2016, 46)
\end{quote}
\textit{Why}, then, is the use of quasi-staticity still so widespread, despite the issues we have discussed? This is because quasi-static processes \textit{do} approximate actual processes in a very concrete way, in terms of size of the driving forces, even though actual processes are \textit{never} exactly quasi-static. We can explicitly see how systems with minuscule driving forces come close to being described by quasi-static processes. For large systems, the time-dependent driving forces are so minuscule relative to the dynamics of the system that we may neglect them for all practical purposes and model them (inexactly!) with quasi-static processes, though we must remember that these systems are \textit{not} undergoing quasi-static processes. Real systems are simply undergoing time-dependent processes which are approximated by quasi-static processes in an inexact fashion. Importantly, Norton shows how we may recover standard thermodynamic results by working purely with time-dependent processes, without getting led astray in pathological cases like processes at the molecular scale.\footnote{See Norton (2016) for discussion.} All this is to say that the exact properties of quasi-staticity are never \textit{essential} to doing thermodynamics. This is good news, since a system bearing the exact properties of quasi-staticity never exists, being contradictory in nature.   

\subsection{Idealization \& de-idealization}

The key lesson of the foregoing is simple but important: we must not confuse properties of the idealization with properties of the target system which approximates said idealization. Quasi-static processes contains properties of change and no-change, but only at the limit of zero driving force. Thankfully, it turns out that actual thermodynamic systems never have exactly those contradictory properties, and thermodynamics does not \textit{essentially} require those properties of quasi-staticity. That's why we may continue using quasi-static processes \textit{qua} approximation.

\2{The broad spirit of this lesson is neither new nor unique to Norton's works. A variety of principles in the philosophy of idealizations converge on this idea. As McMullin (1985) already wrote concerning Galileo's use of idealizations, idealizations\footnote{Specifically, he calls them ``construct idealizations" for reasons extraneous to present discussion.} are typically justified by a clear \textit{de-idealization} process demonstrating how we may remove the simplifications and distortions introduced in the idealization in order to describe realistic systems.\footnote{See also Weisberg (2007).} 

A kindred idea can be found in Earman (2004, 191) when he proposes a \textit{Sound Principle} for interpreting the physical meaning of an idealized model: ``While idealizations are useful and, 
perhaps, even essential to progress in physics, a sound principle of interpretation would seem to be that no effect can be counted as a genuine effect if it disappears when the idealizations are removed." That is, it is to show, in some concrete sense, that the idealization is \textit{dispensable}. Similarly, Fletcher (2020) discusses Duhem's \textit{principle of stability} as an plausible epistemological principle in scientific modeling. Roughly, the principle states that we are justified in inferring, from a model of some phenomenon, conclusions about said phenomenon only if these conclusions remain approximately true of the actual phenomenon when the modelling assumptions only approximately hold.\footnote{For more detailed discussion, see Fletcher (2020).}

While these principles differ in subtle ways, for our purposes they emphasize the same point: to justifiably make inferences about a target system using an idealization, we must have a grasp of what happens when this idealization is removed -- that is, we must have a de-idealization. For instance, to show in the case of phase transitions that an infinite particle number is not needed to make inferences about phase transitions in realistic systems is to show that an arbitrary large number would suffice.\footnote{This is what Wu (2021) proves.} This assumes, of course, that there is \textit{some} way of describing \textit{how} these modelling assumptions approximately hold. If we can show that there exists some de-idealization process for an idealized model, we would also have shown how results from this idealized model approximately hold beyond the model's idealized assumptions, and we can then investigate whether certain effects persist or otherwise, approximately, when the idealization is removed. In my view, then, these principles dovetail in demanding de-idealization in order to justifiably use an idealized model to make inferences about the world. 

Conversely, though, if we \textit{cannot} show how the model approximately holds beyond its idealized assumptions, then we are \textit{not} justified in making conclusions about real-world phenomena from said model. The results from such models would appear to \textit{essentially} require these idealizations, insofar as we cannot show how to obtain them \textit{without} the idealization. 

Returning to the case of thermodynamics discussed above, quasi-static processes are idealizations but they can also be adequately de-idealized in order to describe processes with non-vanishing driving forces. While quasi-static processes \textit{qua} idealizations are useful tools, they are not essential to the description of realistic systems -- there is always a de-idealization procedure available in principle. 

This consideration of whether a property of an idealized system is \textit{essential} to the description of actual systems approximating said idealization has already been much discussed in the literature on the philosophy of thermodynamics. Concerning the physics of phase transitions, Callender (2001), Butterfield (2011), Menon \& Callender (2013) Palacios (2018) and Wu (2021) have all argued, contra Batterman (2002, 2005) that the idealization of a system exactly at the thermodynamic limit is not essential to the understanding of phase transitions, and so the phenomenon of phase transitions does \textit{not} require the actual existence of an idealized system (in this case, a system at the thermodynamic limit with infinite number of particles). Again, a de-idealization procedure is available. This is good news, because real systems undergoing phase transitions are finite and never have the exact (infinite) properties of the idealization. Likewise, as we have seen for the case of quasi-staticity and thermodynamics, quasi-staticity is not essential to thermodynamics. Actual processes are never quasi-static, though they might approximate certain features of quasi-static processes.}

\subsection{Quasi-stationary processes and (the lack of) de-idealization}

So how does the above discussion bear on the issue of black hole evaporation? To start, the same issue with quasi-static processes arises here for quasi-stationary processes: real black holes \textit{cannot} literally be undergoing quasi-stationary processes. Quasi-stationary processes do not refer to actual processes since black holes (or any process) cannot be both time-independent and time-dependent. As such, quasi-stationarity is an idealization. At best, a system may be inexactly described -- \textit{approximated} -- by quasi-stationary processes.

We may then ask: is there a de-idealization procedure by which black holes are approximated by quasi-stationary processes? Furthermore, can this procedure give us something like the global conservation of energy without essentially requiring the idealization of quasi-stationarity? \3{Given that the global conservation of energy is intricately bound up with the (global or asymptotic) symmetries of the spacetime in question, as suggested in \S2, the search for approximate conservation of energy is bound up with the search for \textit{approximate symmetries}, and, in turn, the idea of approximate Killing fields.\footnote{\3{To my knowledge, discussion of symmetries is always bound up with discussion of Killing fields in general relativity. Hence, I take it that approximate symmetries should be understood in terms of approximate Killing fields. Granted, physics is always progressing, and I leave it open that there may be other means of de-idealizing or approximately understanding symmetries and conserved quantities, in which case the reader can understand my project as such: on one very natural way of de-idealizing conserved quantities, via approximate Killing fields, the task of de-idealization faces significant technical troubles. The question is whether these other approaches to de-idealization avoid the worry I raise here, something which can be a fruitful line of research for future work.}} Quasi-stationarity suggests that the spacetime is \textit{almost} stationary, and this suggests, in turn, the need for a procedure for identifying whether an arbitrary spacetime approximates the structure of global \textit{time-like Killing fields} -- approximates a stationary spacetime -- with which one may begin to justify the claim that energy is globally `approximately conserved' in said spacetime.}\footnote{\2{Since, I'll argue, there is no such procedure, the argument here is just a hopeful schematic: if there is a clear way to parametrize how any arbitrary vector field approximates the structure of Killing vector fields, say with a parameter $\epsilon$, we might see how terms vary along this field on the order of $\epsilon$ which can be taken to be arbitrarily small, that is, conserved insofar as $\epsilon$ can be ignored.}} If we can do so, then we might begin to make an argument for black hole evaporation from quasi-stationarity without essentially depending on the idealization of quasi-stationarity. 

It seems to me, however, that there is no clear de-idealization procedure for quasi-stationary processes, in stark contrast to quasi-static processes. Specifically, I will argue that there is a conceptual difficulty in understanding just in what sense a spacetime approximately has a time-like Killing field (or Killing field in general), in stark contrast to the explicit convergence of time-dependent processes, with small enough driving forces, to quasi-static processes, in the case of thermodynamics.

\1{To start off, there is a general worry against the \textit{very idea} of approximate Killing fields, simply because of the nature of general relativity: there is generally no fixed maximally symmetric background spacetime structure against which we have a canonical way of considering deviations from symmetry, i.e. `how close' a spacetime is to being symmetric. For example, there is a natural way in which one can consider deviation from symmetry in Newtonian spacetime (e.g. in terms of asphericity and almost rigid fields). Since Newtonian spacetime is maximally symmetric, flat, and non-dynamical, it provides a fixed background -- akin to a spacetime ruler -- against which we may measure deviations and closeness to symmetry.\footnote{I owe this point to private correspondence with Erik Curiel.} However, there is generally no such fixed background in general relativity with which we can construct such a canonical metric of closeness, save for the special case when the global spacetime is Minkowskian and we do have such a similarly maximally symmetric, flat, and non-dynamical spacetime. Unfortunately, interesting spacetimes are generally not Minkowskian. \textit{Prima facie}, this should warn us against hoping for too much when it comes to seeking approximate Killing fields for general spacetimes.} 

Building on this foundational conceptual worry, and related to this lack of a `spacetime ruler', are three prominent problems related to various attempts at constructing approximate Killing fields. 

\textbf{Problem 1: `closest' does not mean `close'.} These procedures typically provide no clear way to understand how `close' a given vector field is to a Killing field, only that said vector field is, in fact, the closest. This means there is no clear way to understand the deviation from symmetry, and hence evaluate the accuracy of any claims about approximate conservation we might want to make. The fact that the closest In-N-Out burger place to me is in California (as I write from Southeast Asia) is no comfort, for there is no reasonable scale in which it is, in fact, close to me. The situation is worse here, since there isn't even a way to appropriately characterize the `closeness' of a given vector field to a Killing field, unlike a ruler (or its ilk) for the distance between me and the nearest In-N-Out. Without such a notion of `distance', it's not clear to me how one can understand the `approximation' relation, since discussions of approximate symmetry typically employ some such distance function or at least some similarity relation.\footnote{See Rosen (2008) for discussion of this distance function understood as a pseudometric, and Fletcher (2021) for more recent discussion.}

Many extant procedures seek to find the `next best thing' to Killing fields by some generalization of Killing's equation, which, if we recall, is: 

\begin{equation}\label{KE}
    \nabla_{\nu}\xi_\mu + \nabla_{\mu}\xi_\nu = 0
\end{equation}

\noindent As mentioned in \S2, a Killing field satisfies this equation. However, for spacetimes without Killing fields, the equation generally has no nontrivial solutions (Matzner 1968, 1657). Instead, these procedures try to find generalized equations to which a Killing field is but one of many solutions. This is supposed to justify the other solutions as suitable generalizations of Killing fields, insofar as they belong to the same class of solutions. For instance, Beetle \& Wilder (2014) employs an Euler-Lagrange equation of the form 

\begin{equation}
    \Delta_{K} u^{\beta} = \kappa_u u^{\beta} 
\end{equation}

\noindent where what they term the "Killing Laplacian" operator $\Delta_K$ is defined as:

\begin{equation}
    \Delta_K u^\beta := -2 \delta_{\gamma}^{(\beta} g^{\lambda)\nu} \nabla_\lambda \nabla_\nu u^\gamma 
\end{equation}

\noindent Here, $u^{\beta}$ are vector field solutions to the equation with corresponding eigenvalues $\kappa_u$. As they note, Matzner (1968) employs a similar method (though in a slightly different form). This procedure simply defines the `most' approximate Killing field for any given metric to be the vector field solution with the smallest $\kappa_u$ greater than zero. (The solution with a vanishing eigenvalue corresponds to an actual Killing field.) 

However, the problem is that this procedure fails to provide physical meaning to `how close' these generalized fields are to Killing fields. As Matzner (1968, 1657) notes, "we do not have to assume the deviation from symmetry is small" when we are looking for the vector field which best approximates a Killing field. In other words, the most approximate vector field need \textit{not} be close to being a Killing field at all. The approach simply finds a discrete spectrum of eigenvalues (and associated vectors), each increasingly `further' away from being a Killing field. By stipulation, we pick the lowest non-zero eigenvalue and its associated vector field as the most approximate Killing field. Yet, it is not exactly clear in what sense these vector fields are `close' to Killing fields, beyond the fact that these fields become Killing fields when their associated eigenvalues vanish. Along what dimension are these fields becoming `closer' to being Killing fields, and how are we to understand this sort of `distance'? In approximating quasi-static processes with systems possessing small driving forces, we can see how the size of the driving forces dictate the deviation from quasi-staticity. What does the ordering of these generalized vector fields mean here? As Chua \& Callender observes (in a different context, of deriving time's emergence in quantum gravity): ``at the level of pure math, one can ``derive” virtually any equation from any other if one is allowed to assume anything. It makes no sense to say that one equation or quantity is “close” to another absent a metric." (2021, 1176) I think the same is precisely going on here. The approach in question does not provide a physical interpretation of this ordinal ordering of eigenvalues and associated vector fields. The only anchor is the formal fact that the Killing vector field appears as a solution of this more general class of equations, but what does this generalization amount to? Without a convincing story, it remains unclear what it means for a time-dependent physical system, like a black hole, to approximate a spacetime with a Killing field. All we know is that this field is \textit{not} Killing, and that we can define a vector field on it which `best' captures the properties of a Killing field. But is `best' enough? Even if it does generate some vector field that is `close' to Killing, it is still not clear how this is supposed to approximate a Killing field because we are not given a clear understanding of `closeness'. 

A different approach by Cook \& Whiting (2007), adopted in some form by Lovelace et al (2008), has the same problem. This approach compares general vector fields to Killing vector fields on 2-spheres and identifies

\begin{equation}
    S_{ij} S^{ij} = (\nabla_\mu \nabla_\nu v)(\nabla^\mu \nabla^\nu v) - \frac{1}{2}(\nabla^{\alpha}\nabla_{\alpha} v)^2 
\end{equation}

\noindent as a term which vanishes if said vector field satisfies the Killing equation \eqref{KE}.\footnote{Here, $v$ is a scalar field constructed by Cook \& Whiting from the decomposition of a general vector field on 2-spheres.} They then represent how `far' a vector field is from a Killing field by understanding $S_{ij} S^{ij}$ as an `error' term and attempting to extremize $S_{ij} S^{ij}$. However, again, their approach only guarantees that $S_{ij} S^{ij}$ is ``as close to zero as possible" (Cook \& Whiting 2007, 2), but does not provide a clear sense of how close it actually is to a Killing field, and what the meaning of closeness in terms of $S_{ij}$ amounts to physically. They admit that the usefulness of their results for an approximate Killing vector depends on the extent to which a physically meaningful story can be provided for them ``since a Killing vector cannot be produced where one does not exist." (2007, 4) However, they crucially do not provide this story themselves. Yet, a physically meaningful sense in which the $S_{ij}$ term represents `approximation to a Killing field' seems to be exactly what we need here in the present discussion.

Another popular approach discussed by Bona et al (2005) faces the same problem. There is no clear metric for assessing how `close' a given vector field is to being a Killing vector field. Bona et al employs yet another generalization of Killing vector fields via what they call the `almost-Killing' equation. This equation generalizes from the Killing equation by showing that solutions to the Killing equation are also solutions to the almost-Killing equation. Bona et al derives a ``wave equation" of the following form:

\begin{equation}\label{AKE}
    \Box \xi_\mu + R_{\mu\nu}\xi^\nu + (1 - \lambda)\nabla_\mu (\nabla \cdot \xi) = 0
\end{equation}

\noindent where $\Box$ denotes the d'Alembertian operator, $R_{\mu\nu}$ is a Ricci term, and $\lambda$ is a free parameter. They then show that, if $\xi^\alpha$ is a Killing vector, then it satisfies \eqref{AKE}. This provides motivation to take \eqref{AKE} as the generalization of Killing's equation. However, essentially the same problem as the earlier approaches arises here. As Feng et al (2019, 5) point out, vector fields which satisfy the almost-Killing equation need not be in any meaningful sense `close' to being Killing vector fields. Notably, Feng et al observes that any transverse-traceless tensor $Q_{\mu\nu}$ of the following form

\begin{equation}
    Q_{\mu\nu} := \frac{1}{2} (\nabla_\mu \xi_\nu + \nabla_\nu \xi_\mu)
\end{equation}

\noindent satisfies the almost-Killing equation. Of course, when $Q_{\mu\nu} = 0$, that is equivalent to saying that the vector field $\xi_\nu$ satisfies the Killing equation since $Q_{\mu\nu} = 0$ is equivalent to the Killing equation (sans a factor of $\frac{1}{2}$). However, as they point out, $Q_{\mu\nu}$ need \textit{not} vanish, and indeed ``the components of [$Q_{\mu\nu}$] need not be small." (Feng et al 2019, 5) Driving home the point I have been making so far, Feng et al remarks that ``the term “almost Killing” is somewhat of a misnomer" because it's not guaranteed to be almost Killing at all. As with the other approaches I have examined so far, this approach, too, fails to provide a clear sense in which the solutions to these `generalized' Killing fields actually approximate Killing fields. 

In short, the first problem is that these approaches fail to provide a physically compelling story for what it means for this myriad of proposed vector fields to approximate a Killing field, and hence how one might go about de-idealizing the notion of energy conservation. Given the lack of a `spacetime ruler' with which to provide a canonical measure of `distance from symmetry', such a problem might be unsurprising: if there's no canonical measure of `almost-symmetric' in general, then it might be moot to hope for a clear meaning to the claim that something is `almost-Killing'. 

\textbf{Problem 2: no guarantee of time-like Killing fields.} Many of these procedures for obtaining approximate Killing fields do not guarantee that we can obtain an approximate \textit{time-like} Killing field, only that we can obtain \textit{some} approximate Killing fields.\footnote{See, for instance, Matzner (1968), Cook \& Whiting (2007), Lovelace et al (2008) and Beetle \& Wilder (2014).} This means these procedures typically do not tell us how close a spacetime is to having a \textit{time-like} Killing field, but only how close a spacetime is to having \textit{some} Killing field at all. Note, then, that on these procedures a spacetime may not turn out to have anything approximating a time-like Killing vector field at all! These procedures do not necessarily help us de-idealize away from a spacetime with a \textit{time-like} Killing field to a spacetime with some approximate time-like Killing field, but only to a spacetime with \textit{some} Killing field. This, however, is of no help when our goal is to find a de-idealization procedure for the energy conservation of a realistic black hole, in terms of an approximate \textit{time-like} Killing field. What we seek here is a story for how any spacetime approximates one with energy conservation, not just one with any conserved quantity. 

\textbf{Problem 3: overly stringent assumptions.} Many approximation procedures tend to feature strong assumptions which may not be realistic. As mentioned, one can derive almost anything from anything else if we assumed strong enough assumptions, but ``approximations require physical justification." (Chua \& Callender 2021, 1176) For instance, Matzner's (and Beetle \& Wilder's) procedure requires the assumption that spacetime be compact. This is a \textit{very} strong assumption. For one, an arbitrary spacetime can very well be unbounded and infinite, instead of compact, and it is in fact unclear whether our universe is in fact one or the other. Furthermore, a result due to Geroch (1967) suggests that any compact spacetime admits closed timelike curves.\footnote{See e.g. Manchak (2013) for discussion of compactness.} It seems odd that one may only approximate Killing fields in compact spacetimes and not otherwise, given that we may very well live in non-compact spacetimes. As such, these procedures may not even be applicable to the actual world, in terms of which de-idealization would take place. 

Another assumption that frequently shows up is some variant of an assumption of asymptotic flatness. For instance, Matzner's procedure can do without the compactness assumption, if one demands that certain terms of the class of vector fields in consideration vanish ``at infinity", which amounts to some assumption of asymptotic flatness. (Matzner 1968, 1658) It also shows up as an assumption in the almost-Killing equation approach as one way to avoid some scathing problems with the approach. As Feng et al (2019) shows, a Hamiltonian analysis of the almost-Killing equation reveals that the Hamiltonian for this equation is generally unbounded from below. As they explain, ``an unbounded Hamiltonian generally signals the presence of runaway instabilities, which can potentially drive solutions far from the Killing condition." (2019, 2) This means that the almost-Killing equation approach may generate approximate Killing fields which are not close to Killing at all. This returns us to the first problem. 

Feng et al argues that we can avoid this problem for the almost-Killing approach if we assume a vacuum spacetime (for which we can motivate a specific choice of initial conditions, dynamical conditions, and $\lambda$ parameter, hence obtaining a positive-definite Hamiltonian), or if we impose asymptotic flatness as a condition on the spacetime in consideration. (Feng et al 2019, 8) The former does not model any realistic spacetime and does not help us in our de-idealizing. What of the latter? Contrary to the earlier-discussed assumption of compactness, asymptotic flatness is not quite as controversial given its use in much of black hole physics. As such, it seems like a fairly tame assumption. However, I think that any procedure that \textit{requires} the property of asymptotic flatness only passes the buck: we must then provide an account of whether we can de-idealize \textit{that}. In the next section I examine asymptotic flatness in much more detail and argue that, again, we cannot appropriately de-idealize \textit{that} for similar reasons discussed here. 

To sum up, the extant procedures I have examined face three problems. Firstly, and most problematically, there is generally no sense in which a proposed approximate vector field is `close' to a Killing field. Secondly, even \textit{modulo} the first problem, we are not always guaranteed a time-like Killing field. Finally, some of these procedures involve strong and possibly unrealistic assumptions. This complicates the search for a de-idealization procedure away from quasi-stationary spacetime, by preventing us from stating just \textit{how} we are supposed to de-idealize the properties of a Killing field, found only in stationary spacetimes, for non-stationary spacetimes. Granted, I have not proven a negative existential claim here: there \textit{could} very well be a satisfactory procedure in the future. Nevertheless, the above concerns provide strong reasons to be concerned that the argument for black hole evaporation might be lacking in justification, insofar as we cannot explicate what it means for there to be an approximate global conservation law for energy via approximate Killing fields. 

Taken together with the general worry that there is no canonical measure of `deviation from symmetry' above, these problems should pose a significant challenge to anyone looking for an adequate de-idealization procedure which may allow us to relinquish the need for exact conservation of energy via Killing fields by appealing to approximate conservation via approximate Killing fields. (As I discuss in \S6, identical problems return when discussing \textit{asymptotically} Killing fields as well.) 

Barring such a de-idealization procedure, it seems that anyone seeking to employ the quasi-stationary strategy for justifying global energy conservation and motivating black hole evaporation \textit{essentially} needs the idealization of quasi-stationarity, in which a time-dependent evaporating system has time-independent properties, viz. that of having a conserved energy. In other words, one requires an idealized quasi-stationary process which cannot exist. So the argument for black hole evaporation from quasi-stationarity cannot take off yet, and we lack justification at present for using the property of global energy conservation in the \textit{idealized} quasi-stationary system to infer that \textit{actual} black holes, too, have this property. 

\2{Before moving on, I want to emphasize that I don't see my arguments here (or to follow) as a reason to abandon the \textit{general} notion of black hole evaporation altogether, or that this is a reason to think that belief in black hole evaporation is unjustified \textit{period}. Rather, I see this as a challenge for theorists who are keen to use the argument from energy conservation to black hole evaporation to provide a clearer understanding of what approximate (global or, as we'll see, asymptotic) conservation amounts to in the context of general relativity. Given the close ties between symmetries and conserved quantities, the question of what we mean by approximate symmetries must also be tackled.}

\section{Asymptotic flatness and de-idealization}

\3{One might agree with my arguments against quasi-stationarity above, but nevertheless pursue a different strategy for motivating the global conservation of energy. After all, I mentioned in \S2 that global time-like Killing fields are sufficient but not necessary for global conservation of energy; one can also appeal instead to \textit{asymptotic} time-like Killing fields.} \2{For instance, Carroll acknowledges the question of energy conservation but responds that:
\begin{quote}
  ...even in quantum mechanics we have conservation of energy (in the sense, for example, of a conserved ADM mass in an asymptotically flat spacetime). Hence, when Hawking radiation escapes to infinity, we may safely conclude that it will carry energy away from the black hole, which must therefore shrink in mass. (2019, 417)
\end{quote}
Note that ADM (Arnowitt-Deser-Misner) mass is conserved (indeed, can only be defined) in asymptotically flat spacetimes because of the asymptotic symmetries of said spacetime.\footnote{Roughly speaking, we can define ADM mass as the total deviation of the actual spacetime metric from the flat Minkowski metric. More on this in \S6.} Asymptotically flat spacetimes can be considered to have the symmetries of Minkowski spacetime (and can be described by the flat Minkowski metric) \textit{at} spatial infinity, and these symmetries allow for the existence of a time-like Killing field \textit{at} spatial infinity. This then gives us a global conservation law for energy, from which we can, again, motivate black hole evaporation.} 

\subsection{Asymptotic flatness}

If a spacetime is asymptotically flat, then intuitively the 3-metric ``approaches a Euclidean metric sufficiently rapidly at infinity." (Geroch 1972) Then, two metric conditions (among other differentiability conditions) must hold.\footnote{See Geroch (1972, 960) for discussion.} First, outside of some subset of spacetime (where it need not be flat and may contain exotic topological features like singularities), at some arbitrarily large (but finite) distance away from the black hole, the actual metric is equivalent to a Minkowskian metric $h_{\mu\nu}$ with error terms on the order of $r^{-1}$: 

\begin{equation}
    g_{\mu\nu} = h_{\mu\nu} + \mathcal{O}(r^{-1})
\end{equation}

\noindent Second, \textit{at} the limit of spatial infinity:

\begin{equation}
    g_{\mu\nu} \stackrel{r = \infty}{=} h_{\mu\nu}
\end{equation}

\noindent In other words, \textit{at} the limit of spatial infinity, the spacetime is Minkowskian and maximally symmetric. $At$ this limit, we can take the spacetime to possess a time-like Killing field. We can then exploit this symmetry to define a type of global conservation law for energy via the so-called ADM mass mentioned by Carroll. Wallace, too, makes a similar argument:  
\begin{quote}

    Nonetheless we can give powerful arguments for [evaporation]. The most direct is via appeal to Noether's theorem, applied on a sphere surrounding, and far from, the black hole: in that regime, we expect to be able to treat the hole as an approximately-isolated system in a larger region of Minkowski spacetime. So the symmetries of Minkowski spacetime allow us to write a global conservation law and to argue that the sum of the ADM mass-energy of the black hole plus the total energy of the radiation outside the sphere -- which is well defined, since that region is very nearly flat -- should be conserved, and hence that the energy flux through the sphere ought to equal the rate of decrease of the black hole mass. (2018, 11)
\end{quote}

\noindent For context, it is worthwhile to discuss the ADM mass in a little more detail.\footnote{But only a little. The details can be found in the classic Arnowitt, Deser \& Misner (1962).} The ADM approach starts from the Einstein action and derives a Lagrangian from that. Then, by demanding that spacetime be decomposable into time-like and space-like components (i.e. the so-called 3+1 dimensional form), while imposing coordinate conditions at spatial infinity (i.e. that spacetime is flat at $r = \infty$), we arrive at a Hamiltonian formalism and the canonical variables for the target spacetime. Using this formalism, the ADM mass is defined \textit{at} spatial infinity as:

\begin{equation}
    M_{ADM} = \lim_{r\to\infty} \oint (g_{ij, j} - g_{jj, i}) dS_i
\end{equation}

\noindent where $dS_i$ is the two-dimensional surface element \textit{at spatial infinity}, and the indices $i, j$ are over spatial components only. (Arnowitt, Deser and Misner 1962, 16) This integral defines a quantity that can be interpreted as the total energy (of \textit{both} matter and gravity) over a chosen two-sphere, and is zero just when the \textit{entire} spacetime -- including the subset of spacetime we've left out to begin with -- is Minkowskian.\footnote{These are supported by so-called positive mass theorems, first proven by Schoen \& Yau (1979/1981).} Most impressive is the fact that this quantity, while seemingly coordinate-dependent, can be shown to be coordinate-\textit{in}dependent, thus giving us more reason to believe that this quantity has physical meaning. 

Note, however, that this integral is evaluated \textit{at} the limit, and encapsulates Arnowitt, Deser and Misner's key idea that ``the dynamical aspects of the theory are, as expected, to be found in the deviation of the metric from its flat space value." (16, 1962) As they note, ``the basic requirement for an energy to be at all defined is, of course, that space-time become flat at spatial infinity". Only with a flat metric as a standard can we interpret the integral as the deviation from flatness and hence the total amount of energy (i.e. `deviation from vacuum'). The integral is \textit{not} well-defined, nor does it bear a clear physical interpretation, without the existence of flat space at spatial infinity at which it can be evaluated. 

Furthermore, another reason why the ADM mass only has meaning \textit{at} spatial infinity is that it is only \textit{constant}, and hence conserved, at spatial infinity. As Ashtekar \& Hansen (1978) notes, 

\begin{quote}
    Since, in special relativity, the 4-momentum [and hence energy] of a system is intimately intertwined with the group of translations, one might expect the situation to be similar in the present case. This expectation is correct: The 4-momentum emerges as a linear mapping from the space of translations to the reals. Thus, the basic definition of 4-momentum is tied with asymptotic symmetries on [spatial infinity]. (1986, 1556)
\end{quote}

\noindent There is (generalized) time translation symmetry from which we can construct conserved quantities, just like in stationary spacetimes, because of the asymptotically Minkowskian structure at spatial infinity. As Geroch (1972) shows, we can recover the translation group symmetry (and the Poincare group symmetries in general) at spatial infinity, allowing us to introduce asymptotic Killing fields. By considering the conserved quantity along the time-like asymptotic Killing field, we can recover the ADM term straightforwardly. (1972, 965--966) Hence, the symmetries present \textit{at} spatial infinity are what gives the ADM term a clear physical meaning -- that of its conservation over time. 

This conservation law for ADM mass is what motivates defenses of black hole evaporation like those of Carroll and Wallace above. Since ADM mass is conserved for the entire spacetime, any increase in energy at infinity via Hawking radiation must be compensated for by an decrease in energy elsewhere - i.e. the black hole. This then motivates black hole evaporation. \textit{Prima facie}, this argument does not appear to rely on quasi-stationarity, only on the spacetime being asymptotically flat. It seems to avoid the worries I raised above concerning the essential nature of the idealization of quasi-stationarity, and the need for global time-like Killing fields, in motivating a global conservation law for energy. 

While this response seems \textit{prima facie} fruitful, and lines up well with practice, it is nevertheless unsatisfactory for a variety of reasons.

\subsection{Asymptotic flatness, stationarity, and quasi-stationarity}

Let me start first with a minor point. Typical examples of \textit{exactly} asymptotically flat spacetimes like Schwarzschild spacetime, or, more generally, stationary spacetimes are simply non-starters for modelling black hole evaporation, since the dilemma essentially returns. No change to the mass parameter ever occurs in these spacetimes. Furthermore, as Duerr (2021) notes, the general class of \textit{non-}stationary asymptotically flat spacetimes -- the Robinson-Trautman class of metrics describing expanding gravitational waves -- are marred by \textit{naked singularities},\footnote{See e.g. Chruściel 1992 or Podolsky \& Svitek (2005). The aforementioned Vaidya metric belongs to this class of metrics.} and there is a history in physics of an aversion towards singularities due to their seemingly unphysical nature.\footnote{See e.g. Earman 1995, Penrose 1999 for a survey of the orthodoxy.} So it seems, naively, that we are again at a dilemma: stationary asymptotically flat spacetimes do \textit{not} change in time and so black holes in these spacetimes do not evaporate, while non-stationary asymptotically flat spacetimes \textit{do} change in time but are generically unphysical.  

On the face of it, a reply is available: we can rely on quasi-stationarity to model a dynamical black hole changing over time, but \textit{without} relying on it essentially for deriving a global conservation law for energy. We can still model an evaporating black hole with a slowly evolving sequence of stationary asymptotically flat spacetimes (e.g. Schwarzschild spacetimes) by slowly perturbing their mass. However, instead of using \textit{stationarity} to generate the conservation of energy and justify the perturbation of mass in our quasi-stationary process (and hence leading to contradiction), it's the asymptotically flat nature of spacetime which generates an asymptotic conservation law (e.g. of the ADM mass term). It is \textit{this} asymptotic conservation law -- not the one given by stationarity -- that justifies the perturbation of mass and hence the shrinking black hole. Since stationarity \textit{per se} is not essential to our derivation of this asymptotic conservation law, quasi-stationarity -- and the problematic approximate global time-like Killing fields --is not required essentially and hence the problems raised in \S5 are avoided. 

Furthermore, we can ensure the existence of an asymptotic conservation law by noting that every spacetime in the quasi-stationary sequence is asymptotically flat, so that the conservation reasoning always holds from one stage of the quasi-stationary process to another. Since the properties of asymptotic flatness and non-stationarity are not contradictory at each stage, unlike the properties of stationarity and non-stationarity, there is no problem with using quasi-stationarity here as an \textit{approximation} - the idealized, exactly quasi-stationary, process is never required essentially in this argument.

\subsection{Asymptotic flatness as idealization}

So far, so good. The idealization of quasi-stationarity is not required essentially, and so we avoid the contradictions inherent in that idealization. This is fine so long as we are justified in using the idealization of asymptotic flatness in this case. However, it seems to me that we've merely swapped out one conceptually problematic idealization for another. 

To begin, note that deriving the ADM mass required the property of the \textit{existence} of flat spacetime at spatial infinity: the ADM integral is evaluated, and the relevant asymptotic translation symmetries defined, \textit{at} spatial infinity. As Hoefer (2000) and Duerr (2021) have pointed out, however, actual black holes in our universe do \textit{not} live in asymptotically flat spacetimes, since our universe is \textit{not} asymptotically flat. The best model of our universe -- with its predictions (or retrodictions, if one likes) of the Big Bang and the accelerating expansion of the universe -- is an asymptotically \textit{de Sitter} Friedman-Lemaître-Robertson-Walker (FLRW) spacetime with a \textit{positive cosmological constant} $\Lambda$ approaching a (dark energy dominated) de Sitter spacetime in the infinite future.\footnote{See e.g. Fischler et al 2001, Rubin \& Hayden 2016.} 

\textit{Prima facie}, this is a two-fold problem for arguments from energy conservation to black hole evaporation via asymptotic flatness.\footnote{Although Duerr is a little more sanguine when he says that ``it remains to be seen whether the symmetries of de Sitter space allow for a satisfactory formal definition of gravitational energy", I think the situation is a little worse off here.} First, we cannot simply import the results for asymptotically flat spacetimes to describe our world. As Bonga \& Hazboun (2017, 2) notes, ``Since de Sitter spacetime is globally very different from Minkowski spacetime, and most FLRW spacetimes, \textit{none of these calculational tools extend to de Sitter spacetime.}" Second, and worse, there is \textit{no} notion of spatial infinity for asymptotically de Sitter spacetimes, and so the ADM approach cannot even begin here. The de Sitter metric can be written as:

\begin{equation}
    ds^2 = -dt^2 + \cosh^2t d\Omega^2
\end{equation}

\noindent where $d\Omega^2$ is the metric on a unit 3-sphere for 4-dimensional spacetime. As Witten (2001, 1) notes, the spatial sections of such a spacetime, defined by 3-spheres, are spatially compact, and so there is no notion of spatial infinity. In other words, asymptotic flatness does not hold. The relevant asymptotics here are along the \textit{time-like} direction, i.e. future and past infinity. Furthermore, the time-like Killing vectors for de Sitter space do not easily lend themselves to a physical interpretation, since ``there is no asymptotic Killing vector that is globally timelike" here. (Balasubramanian, de Boer \& Minic 2002, 1) As Witten explains: 

\begin{quote}
    In de Sitter space, there is no positive conserved energy. In fact, no matter what generator we pick [...] the corresponding Killing vector field, though perhaps timelike in some region of de Sitter space, is spacelike in some other region. (Witten 2001, 1)

\end{quote}

\noindent Some observers could observe a \textit{negative} conserved energy while others observe a positive conserved energy. Barring exotic accounts of energy, this suggests that there is no physical notion of energy conservation to be defined in this regime. This lines up well with the observation that non-static FLRW spacetimes generically do \textit{not} have a conserved energy.\footnote{See e.g. Mitra (2012).}

In short, then, the assumption of asymptotic flatness, taken literally, cannot hold true of our universe. Arguments for black hole evaporation made in asymptotically flat spacetimes still need to be de-idealized in terms of our actual universe, and for the question of whether black holes in our actual universe in fact evaporate. In particular, actual black holes are \textit{not} literally surrounded by flat spacetime at infinity (and this notion may indeed not even be a possibility at all if our universe is asymptotically de Sitter), yet the existence of flat spacetime at infinity is required for the ADM mass to be defined and for it to be meaningfully conserved in time. If our spacetime is asymptotically de Sitter, then there is seemingly \textit{no} physical global notion of energy conservation to be found. Without a well-defined conservation law for energy, we are still unable to motivate black hole evaporation from Hawking radiation. The present argument from asymptotic flatness has not yet vitiated black hole evaporation. 

\subsection{De-idealizing asymptotic flatness}

The critic might insist that there is no need for asymptotic flatness to be \textit{literally} true of the \textit{entire universe} for an asymptotically flat metric to successfully \textit{approximate} certain systems. Asymptotic flatness may well be an idealization, but an asymptotically flat metric can be (and have been) used effectively to \textit{approximate} isolated systems. If so, we can help ourselves to the existence of flatness at spatial infinity \textit{for all practical purposes}. This then means the existence of an approximately conserved ADM mass is all but guaranteed. I take this to be what Wallace (2018, 11) means in the quote above when he says that far from the black hole, ``we expect to be able to treat the hole as an approximately-isolated system in a larger region of Minkowski spacetime" and that ``the symmetries of Minkowski spacetime allow us to write a global conservation law" for the ADM term since ``that region is very nearly flat".

Agreed: the theme of the previous section is precisely that idealizations can be used fruitfully, \textit{provided} we can provide a clear de-idealization procedure for justifying their use. However, I think the last -- crucial -- step in the above argument, from approximate asymptotic flatness to the existence of an approximately conserved ADM mass is too quick. Even if the asymptotically flat metric successfully approximates an isolated system, it still does not mean that we can help ourselves to \textit{all} of its idealized properties. Recall the distinction we made between approximation and idealization when discussing quasi-staticity and quasi-stationarity. That a process can be approximated by quasi-staticity and quasi-stationarity does \textit{not} mean that it has \textit{all} the properties of quasi-static or quasi-stationary processes. Likewise, when we say that a black hole can be \textit{approximated} by an  asymptotically flat metric, we mean that this metric \textit{inexactly} describes the black hole. Some (perhaps most) properties of this metric describe the black hole to some degree of accuracy, though \textit{not all}. 

It seems to me that the property of flatness at spatial infinity only obtains \textit{at the limit} $r = \infty$, and hence do \textit{not} describe any realistic black holes, since that would mean that the universe is literally empty but for a black hole in its center. Only a (fictitious) black hole with \textit{literally} nothing else in the universe, i.e. flat spacetime at $r = \infty$, is exactly described by the asymptotically flat metric, but this is an \textit{idealization}. 

So the question remains: when a realistic system, such as a black hole or a star, is approximated by an asymptotically flat metric, what properties of this metric can be expected to apply approximately to said system `before the limit' as $r \to \infty$, and which ones only apply when $r = \infty$ and essentially depend on this idealization? The former properties are unproblematic and can be taken to approximately hold for said systems, but the latter can only be said to apply to non-existent fictitious systems like the lone black hole. In Duerr's words, the latter may be thought of as `idle posits' of the idealization which do not refer to properties in the actual system being modelled. Is flatness at spatial infinity something that can be recovered approximately `before the limit', or must it require the limit properties of our fictitious black hole? To put it another way, is there a suitable de-idealization procedure for asymptotic flatness before $r$ reaches infinity? 

To answer this question, we can consider an archetypal use of the asymptotically flat metric: modelling the exterior of a approximately isolated star. This seems to be just the right sort of system to be (and that has been) modelled by an asymptotically flat metric, such as the Schwarzschild metric. (More on \textit{why} they work, in \S6.5.) One way to de-idealize would be to compare the Schwarzschild spacetime with what obtains when we model this star in a (more) realistic spacetime. Only those properties that remain in this more realistic spacetime can be de-idealized, while those properties that only obtain in Schwarzschild spacetime may be said to be non-essential to (more) realistic modelling. One prominent case has been discussed in the literature: when we consider instead a spacetime modelling an approximately isolated system but with a cosmological constant $\Lambda > 0$ -- like the de Sitter spacetime -- the property of asymptotic flatness \textit{never obtains}.\footnote{Koberinski \& Smeenk (2023) discusses similar problems where the quasi-stationary (what they call \textit{adiabatic}) and asymptotically flat assumptions break down, though in the context of critiquing the methods of effective field theories.} 

This suggests that asymptotic flatness is non-essential for modelling realistic stars and is an `idle posit' of the Schwarzschild metric. In this spacetime, the metric does \textit{not} approach a flat one as $r \to \infty$. Furthermore, there is \textit{no} flat Minkowskian region at $r = \infty$. Taking into account any non-zero cosmological constant renders this impossible. As Ashtekar et al (2016) observes, for an isolated system radiating gravitational waves, the asymptotically de Sitter spacetime (which represents the existence of a non-zero cosmological constant) at $r = \infty$ does \textit{not} admit a preferred four-dimensional group of translations unlike a case where $\Lambda = 0$, that is, a case where there is no cosmological constant. In other words, for more realistic spacetimes,\footnote{One example of an asymptotically de Sitter spacetime with a black hole is the Schwarzschild-de Sitter spacetime.} the spacetime do not even begin to approach one with the relevant symmetries for defining something like the ADM mass.\footnote{\1{Rather than use the ADM approach, as the quotes from Carroll and Wallace suggest, we can try to use the Bondi-Sachs approach instead (see e.g. M\"adler \& Winicour 2016). The Bondi-Sachs approach defines a mass term for a system (the `Bondi mass') by ``integrating over an instant of time at null infinity" (Belot 2023, 115), using the symmetries of asymptotically flat spacetimes which only obtain \textit{at} null infinity -- the so-called BMS group -- rather than the Poincare group \textit{at} spatial infinity. The Bondi mass is conserved at null infinity unless there is gravitational radiation, in which case it varies with time depending on the `Bondi news function' which is usually interpreted as energy loss due to gravitational radiation at null infinity. See Fletcher (2024, \S4.4) for some worries with this interpretation.

While my worries about the ADM approach do not directly apply to the Bondi-Sachs approach since the latter is not defined at spatial infinity, it also crucially relies on asymptotic flatness and the BMS symmetries \textit{at} null infinity, and thus faces similar daunting challenges when it comes to de-idealization. On the one hand, the worries} \1{about approximate Killing fields arise (see p. 33 for more discussion) -- in what sense can non-asymptotically-flat spacetimes have approximate asymptotic symmetries and hence approximately asymptotically conserved quantities? On the other hand, to what extent can we retain the physical predictions made using asymptotic flatness, when we remove asymptotic flatness by moving to more realistic spacetimes like asymptotically de Sitter spacetimes? As Saw (2016, 24--25) summarizes, there is significant difficulty in defining the Bondi mass for asymptotically de Sitter spacetimes with a positive cosmological constant. Ashtekar (2017) notes that there are no suitable symmetries at null infinity for asymptotically de Sitter spacetimes which lets us define an analog of the Bondi news term and a possibly-conserved Bondi mass. Furthermore, as Ashtekar \& Dray (1981) noted, while null infinity is fairly unproblematic for \textit{stationary} spacetimes and even some special cases of non-stationary ones, there is no proof that it exists in generic, physically reasonable, non-stationary spacetimes. Finally,  as Belot (2023, Ch. 6, p. 115) points out, ``...already in de Sitter spacetime there are no globally defined time translations -- which renders it impossible to give a general definition of the energy of a system located in a de Sitter spacetime. This problem carries over to the asymptotically de Sitter regime: 
time translations do not appear among the asymptotic symmetries according to any of the standard explications of the notion of an asymptotically de Sitter spacetime." More generally, Belot worries that ``In contrast to their [asymptotically flat] analogs, [asymptotically de Sitter spacetimes] prove to be unsuited for the analysis of conserved quantities and of the flow of
gravitational radiation" (116), and that talk about conserved quantities cannot be brought beyond the asymptotically flat case. I don't think these questions are \textit{a priori} impossible to answer, but as Belot notes, these are challenging open questions, ones I cannot continue to pursue in this paper as it will take us too far afield.}}

One might argue that we could try to find an approximately asymptotically flat spacetime by taking the limit $\Lambda \to 0$, i.e. where we assume that the cosmological constant is vanishingly small and hence approaches an asymptotically flat spacetime. If we can do so, then we can show how there is a clear sense in which the more realistic asymptotically de Sitter spacetime is `almost like a spacetime without a cosmological constant', such as an asymptotically Minkowski -- flat -- spacetime. From there we might then begin to define something like an approximately conserved ADM mass. However, Ashtekar et al (2016, 2--3) show that we will observe discontinuities in the energy term as $\Lambda \to 0$. Koberinski and Smeenk (2023, 12) say the same: ``Minkowski spacetime is qualitatively different from de Sitter spacetime, no matter how small the value of $\Lambda$. Furthermore, the $\Lambda \to 0$ limit is not continuous, as illustrated by the contrast in conformal structure." All this is to say, there is a \textit{qualitative} difference between the asymptotically de Sitter spacetime and the asymptotically flat spacetime. 

Furthermore, these differences are non-trivial and may be empirically significant. In short, they cannot be ignored. As Ashtekar (2017) notes, these factors \textit{do} have physical implications for the system being modelled. For instance, while there is no such limit in asymptotically flat spacetimes, there \textit{is} a novel observationally viable limit on the mass of nonrotating black holes embedded in de Sitter space of $M_{lim} =  \frac{1}{3}\Lambda^{\frac{1}{2}}$,\footnote{See Shibata et al (1994).} which has been corroborated by numerical simulations.\footnote{See e.g. Zilhão et al (2012).} In other words, the properties of the asymptotically flat metric do not even accurately model the system \textit{as} $r \to \infty$, i.e. `before the limit', since it misses out on predictions which the more realistic spacetime provides. As Ashtekar et al (2016, 4) notes: 

\begin{quote}
    A priori, one cannot be certain that the effects of $\Lambda$ would be necessarily negligible because, irrespective of how small its value is, its mere presence introduces several conceptual complications requiring a significant revision of the standard framework. [...] Some of the subtleties associated with a nonzero $\Lambda$ could be important for future detectors such as the Einstein Telescope that will receive signals from well beyond the cosmological radius. They may also be important for the analysis of the very long wavelength radiation produced by the first black holes. (2016, 4)
\end{quote}

\noindent So it seems that the more realistic asymptotically de Sitter spacetime does \textit{not} possess the properties of asymptotic flatness. Nevertheless, it does allow us to recover the predictions made by asymptotically flat metrics, with a caveat:

\begin{quote}
    In the post–de Sitter, first post-Newtonian approximation, it shows that these complications are harmless for binary systems that are the primary targets of the current observatories. (Ashtekar et al 2016, 4)
\end{quote}

\noindent \noindent What this suggests is that we \textit{can} recover the predictions of asymptotically flat spacetimes with more realistic de Sitter spacetimes, \textit{provided we do not go to cosmological scales}. 

In one fell swoop, then, this de-idealization procedure has yielded us two results. Firstly, we learn that the property of asymptotic flatness, that spacetime is flat at $r = \infty$, never obtains in a more realistic setting. Properties which depend on asymptotic flatness, in turn, never obtain either. Furthermore, there is no clear sense in which a spacetime with $\Lambda > 0$ approximates a spacetime with $\Lambda = 0$ by `taking the limit', as we run into discontinuities along the way, suggesting a stark qualitative difference between asymptotically de Sitter spacetime and asymptotically flat spacetime: there is no precise way in which one is `arbitrarily close' to the other (similar to the situation between approximate and exact Killing fields). 

Secondly, we can nevertheless accept that asymptotically flat metrics, with $\Lambda = 0$ have so far been extremely successful. But we can now explain \textit{why} they worked. The de-idealization procedure above suggests that the empirical content generated by these asymptotically flat metrics can be captured by more realistic $\Lambda > 0$ spacetimes. This is simply because these metrics are approximately accurate \textit{up to cosmological scales}. (more on approximate accuracy in $\S6.6$.) However, as we follow the limit of $r \to \infty$, we will find that the model fails to take into account realistic factors affecting the system being modelled, such as $\Lambda > 0$, by assuming that the metric simply approaches a Minkowskian one along this limiting procedure. Once we arrive at scales where $\Lambda$ is relevant, the asymptotically flat metric breaks down. 

Given these two results, we can begin to see how we can rescue all the empirical predictions of asymptotic flat metrics, without actually demanding that asymptotic flatness actually obtains. The more realistic model simply does not have it. Flatness at spatial infinity should therefore be seen as a property of an idealized -- fictitious -- system, as with all the properties depending on flatness at spatial infinity. 

At this point, one might worry: yes, flatness at spatial infinity does not actually obtain in this more realistic model, so there are no symmetries at infinity we can rely on to define a conserved ADM mass. However, we can surely find a region around a system of interest, like a black hole, where spacetime is approximately flat. If we have this, the hope is that we can construct approximate Killing fields along which ADM mass is approximately conserved, \textit{even in this more realistic spacetime}. \2{After all, from eq. 21 we saw that a generically asymptotically flat metric can be described, far away from the source, as a Minkowskian metric plus higher-order terms depending on $r^{-1}$ as it approaches spatial infinity. This is true even if we don't have the exactly Minkowskian metric at spatial infinity. Why can't we just use the asymptotically Minkowskian metric to define asymptotic Killing fields at infinity, but de-idealize by using the higher order terms to define `perturbed' Killing vector fields?

In my view, the problem is that the discussion from \S5.4 applies here as well: there is no good way to exactly understand `approximate' (or `perturbed') Killing fields yet, and essentially the same goes for \textit{asymptotically} approximate Killing fields. We can generate `approximate' Killing fields which are `closest' to being Killing, but this does not actually guarantee that they are `close' to being Killing, and we cannot guarantee that these `approximate' Killing fields are time-like ones with which we can justify approximate energy conservation. This is crucial, because the de-idealization procedure appears to be incomplete otherwise: what we cared about is not the flatness of asymptotic regions \textit{per se}, but how this flatness can be \textit{used to define conserved quantities via its symmetries}. So, even if we can show that the relevant spacetime is approximately Minkowskian in one sense (in terms of the fall-off decreasing with distance), we need to show that this approximation can be used to meaningfully understand approximate conservation of energy. But this requires not only that the spacetime approximately becomes flat, but that this approximately flat spacetime also \textit{approximately has the right symmetries that we care about}, with which we can approximately define global conservation of energy. This is what lets us say that some quantity is approximately conserved -- and the extent to which they are conserved -- at some finite but large distance away from the source. This seems to me to require a grasp of approximate Killing fields, but that is what I am arguing remains out of reach for now. }

Finally, recall that the third problem with approximate Killing fields concerned the reliance of approximating procedures on certain assumptions. Compactness was discussed as an overly stringent assumption, and I left open the discussion on asymptotic flatness. \textit{If} we can assume asymptotic flatness, then we can start with defining certain approximate Killing fields (modulo the above two problems). However, the foregoing discussion suggests that asymptotic flatness cannot obtain given realistic modelling. This suggests that asymptotic flatness, too, is a property of the idealization that does not remain once we attempt to de-idealize, and is likewise an overly stringent assumption. 

It thus seems that the defense from approximate asymptotic flatness to an approximately conserved ADM mass, from which we may deduce black hole evaporation, is unjustified for realistic systems given the present understanding of one justifies idealizations via de-idealizations.  Demanding asymptotic flatness at spatial infinity leads to \textit{inaccuracy} beyond a certain distance scale by failing to take into account well-observed cosmological effects like a nonzero $\Lambda$. De-idealizing a system away from asymptotic flatness, especially for realistic systems, requires the \textit{abandonment} of asymptotic flatness. Furthermore, predictions made with asymptotically flat metrics can be recovered in these more realistic metrics, \textit{without} the property of asymptotic flatness. This suggests that asymptotic flatness is \textit{not} essential to these predictions. Yet, it is precisely asymptotic flatness that is required to define a conservation law for the ADM mass.\footnote{Another intuitive way to see this is that, in Penrose conformal diagrams, spatial infinity is quite literally a point in spacetime \textit{at which} calculations are performed. This suggests that results based on spatial infinity requires the realization of the limit.} We \textit{need} flat spacetime at spatial infinity to compute the ADM integral, and to generate the appropriate asymptotic symmetries. 

If so, the inference of black hole evaporation in our actual world, from asymptotic flatness, lacks justification: this argument requires \textit{essentially} the property of asymptotic flatness, yet exactly this property does not obtain for realistic systems. 

\subsection{Quarantining the idealization}

One might now worry that my argument has gone too far. Isn't asymptotic flatness an ubiquitously used property? For instance, the deflection of light near a gravitating object (or gravitational lensing), as well as the derivation of the famous perihelion precession of Mercury, are cornerstone predictions of general relativity. Numerous measurements have been made which verify these predictions. Importantly, these predictions were historically calculated on an asymptotically flat metric (specifically, the Schwarzschild metric). If asymptotic flatness does not obtain, must one then abandon asymptotic flatness for modelling purposes \textit{completely}? Given how commonplace the use of asymptotic flatness is, and how important some predictions made using asymptotically flat metrics are (say, for confirming general relativity over Newtonian gravity), surely the burden of proof is on us; we appear to be arguing for \textit{too much}. 

I have already argued that we can preserve many predictions made in these asymptotically flat metrics even without asymptotic flatness, as with asymtotically de Sitter metrics up to a certain distance scale. However, given the foundational status of asymptotic flatness, I will attempt to further assuage this worry by considering what \textit{work} the property of asymptotic flatness does for these cornerstone cases, and how that differs from the case of black hole evaporation. 

In my view, asymptotically flat metrics are successful in generating these predictions -- which then allow us to, for example, test and confirm the theory of general relativity over Newtonian gravity and other competitors -- \textit{not} because its asymptotic properties \textit{exactly} hold. We can see this by examining the conceptual role of asymptotic flatness in these benign cases. 

What we want to achieve in these cases, by using asymptotically flat metrics for modelling, is to formally capture a specific property of the target physical phenomena: we care \textit{only} about what general relativity has to say about the gravitational effects of some particular mass on some test particle, and want to consider it as an isolated system, that is, without external influences. By demanding asymptotic flatness, we demand that there is \textit{nothing} in the universe but the gravitating mass; this is what it means to say that spacetime is flat at spatial infinity. Go infinitely far enough away from that one gravitating mass and you'll see nothing. But what justifies the use of this model, in these cases, is \textit{not} the fact that there \textit{really} is nothing at spatial infinity, or even that there \textit{is} an actual realized spatial infinity. Indeed, since what we care about in these cases is the \textit{effect} of the gravitating mass on the test particle, the flat part at spatial infinity is empirically \textit{redundant} insofar as it implies the gravitating mass has \textit{no} effect there. 

What justifies the use of the asymptotically flat metric in these cases, then, is \textit{precisely} the fact that the predictions generated on it hold approximately \textit{even when} the limit property -- that of flatness at spatial infinity -- does \textit{not} hold (as discussed in the previous subsection). How does the de-idealization procedure work in these cases? It is the fact that we can generate predictions from an idealized model of an isolated gravitating system, which we can then de-idealize by seeing how our measurements for said phenomenon matches up \textit{approximately} to these predictions. After all, there \textit{are} external influences beyond the system in consideration: we see other planets, stars, galaxies, and so on. However, predictions produced by this idealized model nevertheless match up to our measurements \textit{to a high degree of precision}, giving us confidence to say that this system can be regarded as \textit{approximately} isolated. Crucially, it is \textit{not} the fact that the predictions \textit{exactly} match our measurements, for which asymptotic flatness will be demanded. 

Recall the spirit of the discussion of de-idealization, Earman's sound principle, and Duhem's principle of stability from \S5.3: even when the idealization does not literally hold true, we must still show in a clear way that the predictions based on this idealization approximately holds in some sense: the predictions must be `close' enough to reality. What we should also say is that the relevant de-idealization also depends on what you are trying to extract from these idealizations. For these benign cases, note that \textit{no talk of approximate Killing fields} is needed to get this de-idealization procedure going for the relevant predictions and modeled behavior. Instead, we can quite straightforwardly show how the prediction holds approximately for the target phenomenon: we have the appropriate \textit{measurements} demonstrating that the predictions hold true of the actual phenomenon to a high degree of precision. The measurements, then, provide justification for continuing to use the idealized asymptotically flat metric, so long as we recognize that asymptotic flatness is in fact an idealization and not \textit{essential} to the use of this model.      

Here's a concrete example: the observational verification and confirmation of relativistic predictions of gravitational lensing over Newtonian predictions. In my view, what mattered in the experimental verification of gravitational lensing was \textit{not} that the calculation \textit{exactly} matched the observations made during, say, the 1919 eclipse, which \textit{would} have made \textit{all} the properties of the metric, including asymptotic flatness, essential for said prediction. Rather, what mattered was two things. 

Firstly, it was the fact that these predictions could be compared to the Newtonian calculation for the \textit{same idealized situation}: that of an isolated point mass gravitating near a ray of light. General relativity predicted that a ray of light will be deflected by 1.75 arcseconds near the Sun, \textit{double} the deflection calculated by Newtonian gravity (0.875 arcseconds). 

Secondly, these predictions were tested against numerous measurements where the criterion for confirmation or verification is \textit{not} the exact matching of predictions to measurements, but only \textit{approximate} consistency in the sense that the predictions fell within the measured results \textit{and the margins of error}. As it turns out, the measurements were consistently closer to the relativistic predictions, \textit{contra} the Newtonian calculations. For instance, the Oxford measurements yielded a deflection of 1.60 ± 0.31 arcseconds (0.91 ± 0.18 times the relativistic predicted value), while the Sobral expedition measurements yielded a deflection of 1.98 ± 0.12 arcseconds (1.13 ± 0.07 times the predicted value). Future measurements continued to be consistent with general relativity, \textit{in the sense that} they are \textit{approximately} the same as relativistic predictions, and differ from Newtonian ones.\footnote{See Will (2015) for a historical account of the measurements of gravitational lensing.} Note the errors in the measurements which confirmed relativistic predictions: the confirmation of general relativity, via gravitational lensing measurements, does \textit{not} require exact matching of predictions to measurements, only that the measurements are \textit{approximately} the same as the predictions. 

In this case, these margins of error provide us with an appropriate de-idealization procedure, by telling us how the predictions do not hold exactly of the target phenomenon, but \textit{only approximately}. In the case of asymptotically flat metrics, we readily see how the measurements \textit{approximate} the idealized predictions made within an asymptotically flat metric. If the idealization \textit{had} held, the measurements would have to hold \textit{exactly} or to \textit{arbitrarily high} degrees of precision. But we never expect it to do so; only that it holds up to a \textit{high degree of precision}, that is to say, \textit{not exactly.}\footnote{The same moral holds for the prediction of the perihelion precession of Mercury: we see how the idealized predictions of general relativity can be de-idealized, when we see how measurements of the perihelion precession of Mercury is accurate \textit{to a high degree of precision}, that is to say, \textit{not exact}.}

It is in this way that such uses of asymptotically flat metrics to generate predictions about astrophysical objects, despite their idealized nature, are justified. Cases like the predictions of gravitational lensing, or the perihelion precession of Mercury, all fall under such uses. Their main goal is to generate relativistic predictions of what is going on in the \textit{non-flat} portion of the asymptotically flat metric.

More generally, the reason why asymptotically flat metrics can still be used to fruitfully model approximately isolated systems is \textit{not} because of the entire global structure of these metrics, but simply a finite patch of it. Specifically, if \S6.4 is right, it is that patch where $r$ is smaller than cosmological scales where $\Lambda$ may be ignored.\footnote{\2{I take it that this is the spirit of Ellis' (2002) proposal to study `finite infinity' and find out ``an effective ‘infinity’ to use" amounts to, ``in discussing boundary conditions for local physical systems", given that real systems are never asymptotically flat and so on.}} To quote Duerr again: 
\begin{quote}
    the working posits of successful asymptotically flat models aren’t their fall-off behaviour at infinity. Rather, they are the right fall-off behaviour \textit{up to cosmological scales}: All empirical content is garnered from the properties of a \textit{finite} patch of an asymptotically flat spacetime. But it’s, of course, the behaviour \textit{at infinity} that is salient of asymptotic flatness. (2021, 27)
\end{quote}
In other words, asymptotic flatness is an \textit{idle posit} of asymptotically flat metrics. No realistic systems being modelled by this family of metrics ever have the limit property of asymptotic flatness. Furthermore, these idealized metrics work well even if asymptotic flatness never \textit{really} holds, because we can extract testable predictions, by just considering a finite patch of this spacetime, with which our measurements are \textit{approximately} consistent. The de-idealization of predictions made within these idealized metrics occurs when we see that our verification of these predictions are always made with the acceptance of measurement errors. That is to say, we never expect these predictions to hold \textit{exactly}, only to a high degree of precision. This then justifies the belief that a system in question so measured, despite never being \textit{truly} isolated, nevertheless can be understood as being \textit{approximately} isolated.

This is in stark contrast to the argument for black hole evaporation, for which (i) we have no measurements to help us in the de-idealization process since the predicted Hawking effect is exceedingly weak, and, more importantly, (ii) essentially relies on the idealized property of asymptotically flat metrics, that is, of \textit{asymptotic flatness}, because we require the asymptotic symmetries for our asymptotically conserved quantities. As we have seen in \S6.1, defining the global conservation of ADM mass \textit{requires} the realization of the asymptotically flat metric by requiring the \textit{existence} of the Minkowskian metric at spatial infinity. De-idealizing away from the asymptotically flat metric leads to a loss of this property for two reasons, as argued for in \S6.4: first, more realistic modelling, taking into account $\Lambda$, cannot demand asymptotic flatness. Second, the definition of ADM mass depends on the presence of asymptotic Killing fields; where there are none, as in realistic situations, de-idealization demands that we produce some approximate notion of these, with which we might perhaps see how ADM mass is `approximately' conserved. However, as I have argued in \S5.4, extant procedures do not generate a satisfying account of how their generated vector fields actually `approximate' Killing fields.

If this is right, then we should really understand the use of asymptotically flat metrics to be justified only for small-scale systems (relative to cosmological scale), and only approximately. Notably, this approximation does \textit{not} include approximate asymptotic flatness. But this already allows us to capture a large swath of actual practice making predictions about astrophysical systems, especially those cornerstone uses which helped us confirm general relativity over Newtonian gravity. 

Two more contemporary uses of asymptotic flatness also merit brief discussion as to how they differ from the use of asymptotic flatness for deriving black hole evaporation via global conservation. Firstly, \2{what about the accretion of matter into a black hole (see e.g. Abramowicz \& Fragile 2013 for a review) and the prediction that black holes grow as a result of infalling matter?\footnote{I thank an anonymous reviewer for suggesting this case.} In that case, don't we \textit{also} use a quasi-stationary sequence of stationary spacetimes (usually Kerr), and justify the perturbation of the mass parameter in the metric in terms of infalling matter? Being a technical topic, I will gloss over the details at a conceptual level. The gist is simply that standard arguments for accretion disks, infalling matter, and black hole growth, do \textit{not} require asymptotic symmetries and global conservation of energy. This allows us to employ the quasi-stationary process as a model for the black hole's mass increase due to accretion, without essentially requiring all the idealized properties of quasi-stationarity or asymptotic flatness (because we do not essentially require time-independence of the metric, globally or asymptotically). This, in turn, sidesteps the worries raised so far. 

It is true that accounts of accretion and infalling matter often appear to invoke gravitational energy or the gravitational potential.\footnote{See also, e.g., Abramovicz et al (1988, 648) which uses the `pseudo-Newtonian' potential as shorthand for modeling ``general relativistic effects in the gravitational field of a black hole" (650).} For instance, Abramovicz \& Fragile starts with a general story for accretion:
\begin{quote}
    In these accretion disks, angular momentum is gradually removed by some presumably (although not necessarily) dissipative process, causing matter to spiral down into the black hole, converting its gravitational energy into heat, and then, by various processes, radiating this  energy. The radiation subsequently leaks through the disk, escapes from its surface, and travels along trajectories curved (in space) by the strong gravity of the black hole, eventually reaching our telescopes. (2013, 6)
\end{quote}
But just as talk about the Newtonian potential $\phi$ is widely understood shorthand for conveniently talking about \textit{curved} spacetime in some appropriate approximation, such parlance about matter `losing gravitational energy', can (and should) also be phrased in terms of matter following trajectories on curved spacetime, past the horizon into the black hole, the derivation of which is entirely local. In other words, the story of accretion and growth can be told entirely without talking about global conservation of energy, because non-localizable gravitational energy does not necessarily need to be invoked even if it is convenient to do so.

What remains to be told is a story for why we are justified in perturbing the mass parameter in any particular stationary spacetime in the quasi-stationary sequence of states, given the infalling matter. This story (told here at a very conceptual level) doesn't necessarily appeal to global conservation of energy, asymptotic or global. As Abramowicz \& Fragile (2013, \S3, \S5.1) summarizes,\footnote{See also Abramowicz \& Straub (2014).} accretion disks are typically given a hydrodynamic description in terms of ordinary conserved (non-gravitational) local quantities such as mass and momentum (defined in terms of the stress-energy tensor $T_{\mu\nu}$), and each point (or `fluid element') of the disk can be attributed a (non-gravitational) stress-energy tensor, which results in models such as the `Polish Doughnut' in the simplest case of a perfect fluid on curved spacetime.\footnote{See Abramowicz et al (1988).} Then, given some (finite!) patch of a black hole spacetime such as the Kerr spacetime (specifically, the region in, on, and around the horizon), the question of infalling matter simply becomes the question of which trajectories will carry these fluid elements, and local quantities on them, past some volume element of the horizon into the black hole, and which will not. But it follows from \textit{local} conservation of $T_{\mu\nu}$ that we can track how much mass/energy has been transported across some volume element of the horizon along any particular trajectory. In turn, this gives us a justification for perturbing the mass parameter of the stationary spacetime (and hence for using the quasi-stationary process). Regardless of how much gravitational energy there is, it's simply the case that this local transport of mass into the horizon \textit{will} increase the total mass inside the black hole, this time \textit{without} appealing to global conservation, the time-independence of the metric, and the existence of global time-like Killing fields, avoiding the problems raised for the argument for black hole evaporation from global conservation. As such, black hole accretion need not face the same worries that I've raised for the family of arguments for black hole evaporation via global conservation of energy.

The key difference between arguments about accretion and the sort of arguments for black hole evaporation I've considered here, via global energy conservation, is that the story about accretion and infalling matter can be an entirely \textit{local} affair. However, in the family of arguments I've been considering, Hawking radiation is derived using \textit{global} spacetime structure, for instance by comparing global quantum fields at past and future null infinity in a black hole spacetime, and is therefore a property of the \textit{entire} horizon rather than any particular localized point on the horizon. Likewise, then, black hole evaporation becomes a global affair, and requires considerations about global conservation of energy as I have emphasized so far. 

Now, seemingly \textit{local} arguments from vacuum polarization of the expected stress-energy tensor (rather than global conservation of energy) for black hole evaporation, mentioned in fn. 6, may also employ similar reasoning to the case of accretion, insofar as they work primarily with the local (expected) vacuum stress-energy tensor and its behavior near the horizon at each point, rather than with global spacetime structure. While this other family of arguments is not my focus here, it could provide an alternative route to black hole evaporation which may be less problematic -- and hence much more attractive -- given the discussion of this paper so far. Notably, this might mean we ought to give up the sort of global argument often provided in the literature -- including in Hawking's original arguments -- and adopt such local arguments instead, unless we can find a way to de-idealize asymptotic flatness or quasi-stationarity.} 

\1{Finally, I consider a case that is less open-and-shut. What about standard derivations of gravitational waves, which also appear to rely on asymptotic flatness in crucial spots?\footnote{I thank an anonymous reviewer for suggesting this.} Getting into the controversial details of gravitational waves and its energy will, again, take us too far afield (see e.g. Hoefer 2000, Lam 2011, Duerr 2019, Cartwright et al 2023 Ch. 7), so I will make some broad observations. I think we can distinguish two questions here. One, can standard derivations for the \textit{existence} of gravitational waves be de-idealized? Two, can standard derivations that gravitational waves \textit{carry energy} be de-idealized? 

To the first question, I think the answer is \textit{yes}. Standard derivations of gravitational waves appeal to linearized gravity in the weak-field regime, for which the metric can (again) be decomposed into a flat Minkowskian metric plus some perturbation term. By imposing the appropriate gauge conditions one can derive solutions to this perturbation term which are solutions to a wave equation. The details are much more than I can provide here, but the crux is that what matters is the \textit{fall-off} -- the separation of the metric into two parts, one of which can be seen as wave-like under specific conditions -- as we \textit{approach} spatial infinity (the first condition of asymptotic flatness, eq. 21), rather than the Minkowskian properties \textit{at} infinity, i.e. the derivation does not essentially require the existence of a flat spatial infinity (i.e. the second condition of asymptotic flatness, eq. 22). So, at the very least, my arguments against asymptotic flatness and its resistance to de-idealization doesn't complicate issues for the existence of gravitational waves per se. 

But what about the second question? We can agree that gravitational waves \textit{exist}, but it's a further question -- one that has a storied history (see Cartwright et al 2023, Ch. 7) -- as to whether (and how) it actually carries energy away from its source. Here I think that the problems with de-idealizing asymptotic flatness return for reasons already discussed in the literature. Given the technical and broad nature of \textit{that} debate, I will only skim the discussion here.\footnote{\1{For arguments against the claim that gravitational waves carry energy, see Hoefer (2000), Duerr (2019), Fletcher (2024, Ch. 4). See Curiel (2019b) for a proof, under plausible conditions, that there can be no gravitational stress-energy tensor, suggesting a difference in kind between gravitational energy and ordinary forms of energy represented by stress-energy tensors (e.g. energy of the electromagnetic field). For a recent counterpoint, see Gomes \& Rovelli (2023).}} Roughly, the claim is that gravitational waves carry away energy from their source, e.g., rotating binary pulsars like the Hulse-Taylor binary pulsar. It is commonly said -- given what Duerr (2019) calls the `standard interpretation' -- that their orbits observably decay (which won Hulse and Taylor the 1993 Nobel prize) and that this ``can be seen as an indirect proof of the existence of gravitational waves" (Nobel Prize Committee 1993). The idea is that the best explanation is that gravitational waves carry away energy leading to the decay of their orbits. This, of course, relies on some principle of energy conservation. Furthermore, since there is no local notion of the energy of gravitational waves, this principle must be about \textit{global} conservation of energy -- which requires a global or asymptotic symmetry as I have argued (for gravitational radiation, the Bondi-Sachs formalism, problematized in fn. 45, rather than the ADM formalism is often used). Insofar as there is no de-idealization procedure for how realistic systems can approximately have such symmetries, we are not yet justified to use such principles. One way out is to adopt Duerr's (2019, \S3.4) \textit{dynamical} interpretation of binary pulsars, which explains the orbital decay of binary pulsars simply in terms of the relativistic equations of motion and the Einstein field equations, \textit{without} appealing in any way to gravitational waves or energy, sidestepping the worries with de-idealizing the relevant asymptotic symmetries. If Duerr is correct, then the use of gravitational energy, global conservation of energy, and all the idealizations required to make sense of those notions, though useful, are not \textit{essential} for explaining the behavior of gravitational waves and the orbital decay of binary pulsars. 

I admit that the case of gravitational radiation is less of an open-and-shut case than I would have liked, and that there remains work to be done to sort out the situation here. Perhaps advances in this debate can even help us out in de-idealizing the case of black hole evaporation -- I welcome such developments. This is part of the point of the paper: there remains much work to be done to understand how and why we are allowed to use the varied idealizations in general relativity, and when these idealizations break down, work that can be pursued by philosophers and physicists alike.}

I hope to have shown here that we \textit{can} quarantine the argument against asymptotic flatness from most uses of asymptotic flat metrics, because many of those uses do \textit{not} essentially require asymptotic flatness -- the existence of a flat spatial infinity -- for the generation of testable predictions. However, insofar as the family of arguments for black hole evaporation currently in question requires the premise of global energy conservation, \textit{that} argument does not hold for realistic systems. 

\section{Concluding Remarks}
 
Recently, Read (2020, 17) argues \textit{contra} Hoefer (2000) (who also argues that our universe is not asymptotically flat, so has no conserved ADM mass) that:

\begin{quote}
   every theory of physics is an idealization and does not ‘apply to the actual world’ in this strong sense. So, Hoefer’s objection [...] seems at the same time to apply to an unacceptably broad class of physical laws and theories.
\end{quote}
If Read is right, then it seems that I am asking for the impossible from physics. However, I see it as a mere deflection, not a confrontation, of the underlying problem. I agree with Read that idealizations \textit{qua} idealizations are unproblematic. What I find problematic -- in line with Norton -- is the assumption that the systems approximating those idealizations inherit \textit{all} the desirable properties of those idealizations. This is something which must be investigated on a case-by-case basis by constructing appropriate de-idealization procedures, and what is already done by some like Ashtekar and his colleagues. 

On that front, it seems that we are, at present, lacking good justification for buying one prominent family of arguments for black hole evaporation, those which employ the global conservation of energy. Two idealizations appear to provide refuge from the n\"aive dilemma I raised in \S4: quasi-stationarity and asymptotic flatness. They are supposed to justify how time-dependent spacetimes can approximately have symmetries required for approximate global conservation of energy either by appealing to global or asymptotic Killing fields. However, both appear to be unjustified for realistic systems once we attempt to de-idealize these highly idealized assumptions. So we are still left, again, with the naïve dilemma for the argument for black hole evaporation from conservation of energy. I reiterate that I have not argued that there could \textit{not} be such a justification; I hope that this paper is seen, on a positive note, as a challenge for theorists to provide some such justification or clarification. 

For philosophers, I hope to have shown that black hole physics remains a rich wellspring for philosophical debate. Black hole thermodynamics (and black hole physics in general) is a field where many approximations are made, seemingly without issue. However, as I have shown, a closer look may reveal some conceptual puzzles, such as that of quasi-stationarity, as well as empirical concerns, such as that of asymptotic flatness. 

My discussion also relates -- and adds -- to well-known discussions about idealizations versus approximations, by analyzing whether the properties of idealizations like quasi-stationarity transfer over to the systems approximating such idealizations, and whether black hole evaporation is a phenomenon that essentially requires the idealization of quasi-stationarity. As I have argued, careful handling of idealizations and approximations reveals that black hole evaporation lacks justification. 

The present discussion also connects with the literature surrounding the broader problem of defining a suitable notion of energy in general relativity. I hope to have shown how these foundational discussions can be of import to contemporary physics. 

I foresee much more to unpack on this front, and hence more work to be done. Here, I have only explored the troubles with defining a global conservation law via global time-like Killing fields, and the conceptual difficulties with finding suitable approximate descriptions of a globally conserved energy term via quasi-stationarity or asymptotic flatness for realistic systems. Recently, however, there has been some focus on so-called quasi-local alternative definitions for energy: two notable proposals are the pseudotensor approach and the Brown-York quasi-local approach. The problems with the quasi-pseudotensor approach are already well-known both in physics and philosophy. The Brown-York tensor approach, which defines a quasi-local energy via an integral on the 2-surface of a 3-volume in spacetime, appears more fruitful. McGrath (2012) and De Haro (2022) have both argued that this quasi-local approach is conceptually superior to the global and pseudotensor approaches to defining energy. However, it appears that the Brown-York quantity still requires the existence of at least a global time-like Killing field in order for it to be conserved. (Szabados 2009, 85) Since spacetimes with time-like Killing fields are stationary, the dilemma I have presented here may threaten to return.\footnote{De Haro (2022) does note that the condition can be weakened to the requirement of merely \textit{conformal} Killing fields for spacetimes with a positive $\Lambda$, which merits some study.} Interestingly, Beetle \& Wilder (2014, 2--3) notes that the Brown-York approach depends on the ability to define a 2-sphere $S$ which is ``fixed geometrically by features of the surrounding spacetime", and thus is ``on their surest footing when the intrinsic geometry on $S$ admits an axial Killing field." Notably, they continue that when there are no Killing fields, as is the case for generic spacetimes:
\begin{quote}
    when $S$ is not axially symmetric, one must specify how to construct a similarly geometrically preferred vector field $\xi^\alpha$ from an arbitrary 2-sphere metric. Moreover, this vector field must reduce to the axial Killing field whenever one exists. A natural way to do this is to seek a best approximation, in some prescribed sense, to a Killing field on $S$.
\end{quote} 
In other words, it seems that the Brown-York approach, if it is to take off, must \textit{also} be justified by an appropriate de-idealization procedure. If so, it may well be the case that the problems with de-idealization I have pointed out here might infect these quasi-local approaches. 

Finally, one might think that my argument only works because of a highly stringent standard for what counts as justification of an idealization, that of de-idealization. I have two things to say here: first, this notion is popular in philosophy of science, and is implicitly at work in discussions about justifying other idealizations, for instance, the thermodynamic limit (e.g. Wu 2021) or the ergodic hypothesis (e.g. Vranas 1998) in thermodynamics and statistical mechanics. Second, I agree that this is a high standard, but there is not much that has been done to `de-idealize' the notion of de-idealization by understanding other ways in which idealized models can be tethered to the world without resorting to `taking the limit' within or between models of physics. On this more relaxed view, it might just be the case that we \textit{can} justify the use of certain idealizations for black holes. However, no such account exists at the moment. This is, in fact, something I am working on at the moment, but I leave it for future work. 



\bibliography{citations}
\nocite{*}
\end{document}